\documentclass[fleqn,usenatbib]{mnras}

\usepackage[T1]{fontenc}
\usepackage{ae,aecompl}

\usepackage{graphicx}	
\usepackage{amsmath}	
\usepackage{amssymb}	
\usepackage{tikz}
\usepackage{newtxtext,newtxmath}

\newcommand{\hmpc}{\,$h$\,Mpc$^{-1}$}
\newcommand{\mpch}{\,Mpc\,$h^{-1}$}
\newcommand{\msunh}{\,${\rm M}_\odot/h$\,}
\newcommand{\arepo}{\textsc{Arepo}}
\newcommand{\music}{\textsc{MUSIC}}

\title[ETHOS haloes: abundance and inner structure at high$-$z]{The halo mass function and inner structure of ETHOS haloes at high redshift}

\author[S. Bohr et al.]{Sebastian Bohr$^{1}$\thanks{E-mail: \href{mailto:seb21@hi.is}{seb21@hi.is}}, Jes{\'u}s Zavala$^{1}$, Francis-Yan Cyr-Racine$^{2}$ and Mark Vogelsberger$^{3}$
\\
$^{1}$Centre for Astrophysics and Cosmology, Science Institute, University of Iceland, Dunhagi 5, 107 Reykjavik, Iceland \\
$^{2}$Department of Physics and Astronomy, University of New Mexico, 210 Yale Blvd NE, Albuquerque, NM 87106, USA\\
$^{3}$Department of Physics, Massachusetts Institute of Technology, 77 Massachusetts Avenue, Cambridge, MA 02139, USA
}

\date{Accepted XXX. Received YYY; in original form ZZZ}

\pubyear{2021}

\begin{document}
\label{firstpage}
\pagerange{\pageref{firstpage}--\pageref{lastpage}}
\maketitle

\begin{abstract}
We study the halo mass function and inner halo structure at high redshifts ($z\geq5$) for a suite of simulations within the structure formation ETHOS framework. Scenarios such as cold dark matter (CDM), thermal warm dark matter (WDM), and dark acoustic oscillations (DAO) of various strengths are contained in ETHOS with just two parameters $h_{\rm peak}$ and $k_{\rm peak}$, the amplitude and scale of the first DAO peak. The Extended Press-Schechter (EPS) formalism with a smooth-$k$ filter is able to predict the cut-off in the halo mass function created by the suppression of small scale power in ETHOS models (controlled by $k_{\rm peak}$), as well as the slope at small masses that is dependent on $h_{\rm peak}$. Interestingly, we find that DAOs introduce a localized feature in the mass distribution of haloes, resulting in a mass function that is distinct in shape compared to either CDM or WDM. We find that the halo density profiles of {\it all} ETHOS models are well described by the NFW profile, with a concentration that is lower than in the CDM case in a way that is regulated by $k_{\rm peak}$.  
We show that the concentration-mass relation for DAO models can be well approximated 
by the mass assembly model based on the extended Press-Schechter theory, which has been proposed for CDM and WDM elsewhere. Our results can be used to perform inexpensive calculations of the halo mass function and concentration-mass relation within the ETHOS parametrization without the need of $N-$body simulations.
\end{abstract}




\section{Introduction}
\label{sec:introduction}

A majority of the matter content of the Universe is made up by dark matter (DM), which is therefore a crucial ingredient in cosmological structure formation. A likely explanation for DM is that is made of yet undiscovered particle(s), whose nature remains a mystery. A prominent assumption within the particle hypothesis is that taken by the Cold Dark Matter (CDM) model, which in essence states that the only DM interaction relevant for structure formation is gravity. CDM has been established as the standard paradigm for structure formation since it has been shown to be consistent with the observed structure of the Universe on large scales \citep[e.g.][]{Springel2005}. However, the CDM model remains challenged on smaller (galactic) scales in various ways: (i) the underabundance of low-mass galaxies (either satellites or in the field) \citep{Klypin1999,Moore1999,Zavala2009,Papastergis2011,Klypin2015}, 
(ii) the core-cusp problem in low-surface brightness galaxies and possibly in dwarf spheroidals \citep{deBlok1997,Walker2011},
(iii) the "too-big-to-fail problem" \citep{Boylan-Kolchin2011,Papastergis2015}, (iv) the plane of satellites problem \citep{Pawlowski2013}, and (v) the diversity problem of rotation curves in dwarf galaxies \citep{Oman2015}. For recent reviews on the CDM challenges and plausible solutions see \citet{Bullock2017} and \citet{Zavala2019}.

A possible approach to address these potential issues 
is to invoke additional DM physics, i.e., to consider departures from the CDM hypothesis that change its predictions on small scales while leaving the large scale behaviour intact. A novel framework (ETHOS) has been proposed to incorporate new DM physics into structure formation theory, connecting a broad range of DM particle physics to effective parameters that characterize structure formation in the linear regime \citep[][]{Cyr-Racine2016,Vogelsberger2016}, and further to effective parameters that capture the behaviour of different DM models in the non-linear regime \citep[][]{Bohr2020}. 
The new parametrization introduced in \citet{Bohr2020} is based on describing dark acoustic oscillations (DAOs) in the linear power spectrum. The two physically motivated parameters $h_{\rm peak}$ and $k_{\rm peak}$, the amplitude and scale of the first DAO peak, respectively, suffice to describe the linear power spectrum for DM models from WDM ($h_{\rm peak}=0$) over weak DAOs (wDAO; $h_{\rm peak}\sim0.2$, like those in \citealt{Vogelsberger2016}) to strong DAOs (sDAO; $h_{\rm peak}\sim1$, like those in \citealt{Bose2019}). In \citet{Bohr2020}, it was shown the parameter space of DM ETHOS models in the ($k_{\rm peak},h_{\rm peak}$) can be divided clearly in distinct structure formation regions (CDM-like, WDM-like and DAO-like). When this division is done according to the non-linear power spectrum at high redshift, only a small region of the parameter space still displays distinct DAO features by $z=5$. This DAO region can be augmented if the halo mass function is used instead as a measure to classify the models; \citet{Bohr2020} found that
the halo mass function is especially sensitive to the presence of DAO features in the linear power spectrum.

In this work, we apply the Extended Press-Schechter (EPS) formalism \citep[][]{PS1974,Bond1991,Sheth1999} to a wide range of ETHOS models, which has not been done before broadly (\citealt{Sameie2019} applied this formalism to the small subset of wDAO ETHOS models in \citet{Vogelsberger2016}), and tweak it to accurately represent the simulated halo mass function. The use of this formalism offers a quick way to compute the halo mass function without the need to run dedicated and computationally expensive $N$-body simulations. 

The non-linear power spectrum at small scales depends both on the halo mass function and the inner structure of DM haloes, both of which are affected by the DM nature. In particular cut-offs and additional features in the linear power spectrum due to new DM physics have been shown to affect not only the abundance of DM haloes, but also their inner density profile. For instance, for WDM it has been shown that DM haloes still follow a NFW density profile, but with lower concentration for small haloes relative to CDM \citep[see e.g.][]{Lovell2014,Ludlow2016}. On the other hand, for DAO models, it has been shown that haloes become overall less centrally dense due to the suppression of power at small scales \citep[see e.g.][]{Buckley2014,Vogelsberger2016}. However, the inner halo properties of DAO models have not been studied in detail, in terms of their dependence on the scale and amplitude of the DAOs. This is something we pursue in this work by looking at the halo concentration in ETHOS models and attempting to predict its behaviour using the model of \citet[][]{Ludlow2016} coupled with the EPS formalism.

Our work focuses on the high redshift regime ($z\geq5$) to test the limits of the EPS formalism and the concentration model of \citet[][]{Ludlow2016} for ETHOS models. The high redshift regime has been shown to be a promising one to probe and distinguish different ETHOS models \citep[e.g. see][for predictions for the 21-cm signal]{Munoz2020} and it is therefore important to test the validity of analytical approaches such as EPS. Our work is also motivated by a lack of previous work studying the inner structure of haloes at high redshift for DAO models.

This paper is organized as follows. In Section~\ref{sec:method}, we shortly summarize the setup of the simulations used in this work. Section~\ref{sec:hmf} covers the EPS formalism for the halo mass function and its application to our set of ETHOS simulations. In Section~\ref{sec:inner}, the inner halo structure is studied by looking at the concentration parameter of DM haloes. Finally, our conclusions are given in Section~\ref{sec:conclusion}.

\section{Simulations}
\label{sec:method}

In this work, we use the cosmological DM-only $N$-body simulations that were described in detail in \citet{Bohr2020}; they were performed with the code \arepo~\citealt{Springel2010} from initial conditions generated with \music~\citealt{Hahn2011}. All simulations use the cosmological parameters $\Omega_{\rm m}=0.31069$, $\Omega_\Lambda=0.68931$, $H_0=67.5\,{\rm km/s/Mpc}$, $n_{\rm s} = 0.9653$ and $\sigma_8=0.815$, where $\Omega_{\rm m}$ and $\Omega_\Lambda$ are the matter and cosmological constant contributions to the matter-energy density of the Universe today, respectively, $H_0$ is today's Hubble constant, $n_{\rm s}$ is the spectral index, and $\sigma_8$ is the mass variance on 8\mpch scales. The high resolution region of the simulations has a comoving smoothing length of $\epsilon=0.2\,{\rm ckpc/h}$ and a particle mass of $8\times 10^4$\msunh.

\begin{figure}
    \centering
    \includegraphics[width=\columnwidth]{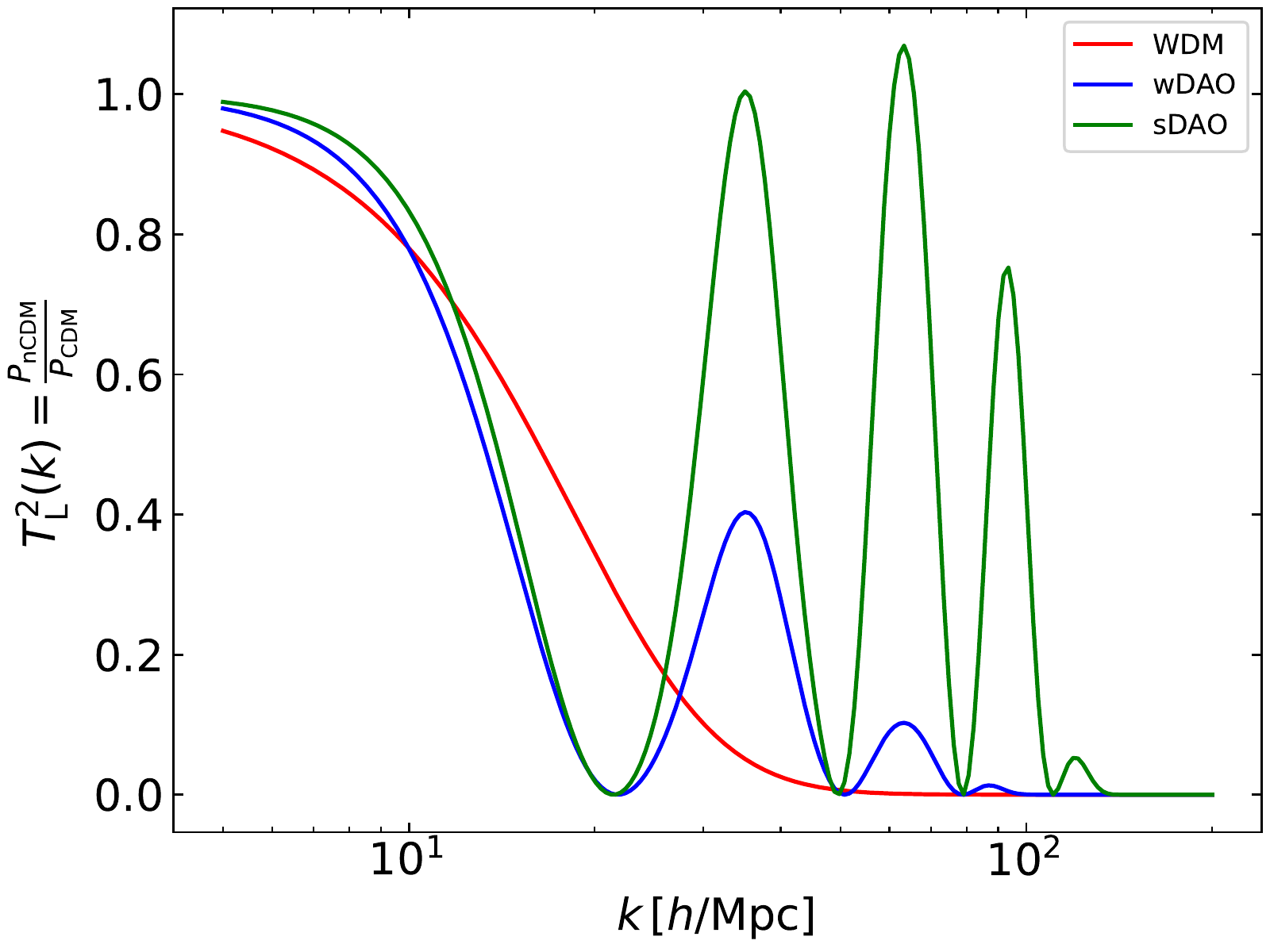}
    \caption{Initial linear transfer functions $T^2_{\rm L}(k)$ for examples of WDM (red), wDAO (blue) and sDAO (green) models. All three models have identical $k_{\rm peak} = 35$\hmpc}
    \label{fig:ICTransfer}
\end{figure}

The suite of simulations covers CDM ($k_{\rm peak}\rightarrow\infty$) and WDM-like models ($h_{\rm peak}=0$) in a wide range of cut-off scales ($k_{\rm peak}=35-300$\hmpc; equivalent to WDM masses $m_\chi \approx 1.6-11\,{\rm keV}$). The suite covers a range of DAO models from weak DAOs with $h_{\rm peak}=0.2-0.6$ to strong DAOs with $h_{\rm peak}=0.8-1$ \citep[for the effect of sDAO features on the Lyman-$\alpha$ forest, see][]{Bose2019} and DAO scales of $k_{\rm peak}=35-300$\hmpc. We note that some of the WDM models explored here are already in tension with current constraints on the non-linear power spectrum from Lyman-$\alpha$ observations, e.g. the allowed WDM masses $m_{\rm WDM} > 3.6\,{\rm keV}$ \citep{Murgia2018} would correspond to $k_{\rm peak}\gtrsim85$\hmpc in the limit $h_{\rm peak}\rightarrow0$, which is the WDM limit in our parametrization. Since wDAO models show degeneracies with WDM in the matter power spectrum, the wDAO models with $k_{\rm peak}\lesssim65$\hmpc are also ruled out by the same observations in a way that is predicted in  \citet{Bohr2020} (see Fig. 10 therein). For sDAO models, there is a single simulation including baryonic physics that explores the impact of sDAO features in the Lyman-$\alpha$ forest 1D flux spectrum \citep{Bose2019}, and given their quite distinct behaviour relative to WDM, a detailed analysis is needed to properly set constraints in sDAO models. Overall, only a few of the models in the simulation suite we use can be considered as being ruled out by current observations, but we nevertheless include them here for illustrative purposes of the extreme behaviour in the wDAO and WDM regimes at low $k_{\rm peak}$ and low $h_{\rm peak}$.

Figure~\ref{fig:ICTransfer} shows the linear transfer function of examples of WDM, wDAO and sDAO models with identical $k_{\rm peak}$. Physically, the DAO models are characterised by the sound horizon scale at the time of DM-DR decoupling, which essentially sets the scale of the first DAO peak, $k_{\rm peak}$, while the amplitude of this peak is determined by the timescale of the DM-DR decoupling relative to the Hubble rate (see Section 3.1 in \citealt{Bohr2020}), which is what sets the difference between the wDAO and sDAO regimes. 
A faster decoupling timescale 
leads to a fast transition from the tightly coupled regime to the decoupled regime and the DM power spectrum does not get damped significantly (sDAOs). For larger decoupling timescales, there is a slow transition between these regimes, with
the extended period of the weakly coupled regime dampening the DAOs significantly (wDAOs). 

Finally, we remark that ETHOS models self-consistently contain astrophysically relevant self-interacting cross sections, which can impact the inner structure of DM haloes \citep[see][]{Cyr-Racine2016,Vogelsberger2016}. However, as in \citet{Bohr2020}, in this paper we only consider the effect of the primordial suppression of the matter power spectrum and leave a study of the effect of possible DM self-interactions for future work. We do this for two reasons. First, we want to cleanly separate the effects of the primordial suppression and DM self-interactions. Second, we expect self-interactions to be more relevant at lower redshifts than studied here ($z \gtrsim 5$; see e.g \citealt{Vogelsberger2014b}). From the simulations, the haloes were constructed using FOF and SUBFIND algorithms included in \arepo~with a particle number limit of 32. For more details on the simulations, see \citet{Bohr2020}. 

\section{Halo mass function in ETHOS haloes at high redshift}\label{sec:hmf}

For the effect of different ETHOS models on haloes, we first look at their abundance as measured by the halo mass function. For the halo mass function, we do not include subhaloes, but purely main haloes.

\subsection{Extended Press-Schechter formalism}
The halo mass function can be modelled from the linear power spectrum using variants of the Press-Schechter formalism \citep{PS1974,Bond1991,Sheth1999,Sheth2001} 
The following is a brief summary of the key equations in the variant we will use.

Regions with a characteristic size $R$ corresponding to a mean mass scale:
\begin{equation}
M = \frac{4\pi}{3}\bar\rho_m R^3, \label{eq:M}
\end{equation}
where $\bar{\rho}_m=\Omega_m\rho_c$ is the mean matter density ($\rho_c$ is the critical density of the Universe), have a smoothed density field $\delta_M$:
\begin{equation}
    \delta_M\equiv\delta(\vec{x};R)=\int\delta(\vec{x}')W_R(\vec{x}-\vec{x}';R)d^3\vec{x}\label{smooth_dens}
\end{equation}
where $W_R$ is a window or filter function properly normalised, and $\delta(\vec{x})$ is the matter density contrast. The (linear) mass variance is the most relevant statistical quantity of the smoothed density field in the Press-Schechter formalism, and it is given by:
\begin{equation}
    \sigma^2(R) = \frac{1}{2\pi^2} \int_0^\infty dk k^2 P(k) \widetilde{W}_R^2(k) \label{eq:sigma}
\end{equation}
where $P(k)$ is the linear power spectrum and $\widetilde{W}_R(k)$ is the Fourier transform of the window function in Eq.~\eqref{smooth_dens}. 

In the Extended Press-Schechter (EPS) formalism, it is then argued that the comoving number density $n(M)$ of collapsed haloes of mass $M$ (Eq.~\ref{eq:M}) is given by:
\begin{equation}
    \frac{dn}{d\ln M} = -\frac{1}{2} \bar{\rho}_m \frac{f}{\sigma^2} \frac{d\sigma^2}{dM} \label{eq:hmf}
\end{equation}
where $f(\nu)$ is the so-called first crossing distribution (or multiplicity function) within the ellipsoidal collapse model (see \citealt{Sheth2001}):
\begin{equation}
    f(\nu) = A\sqrt{\frac{2q\nu}{\pi}}(1+(q\nu)^{-p})\exp\left(-\frac{q\nu}{2}\right)\label{eq:ellipsoidal}
\end{equation}
where $p=0.3$, $q=1$, and we fit $A$ with our simulations, while $\nu$ is defined in terms of the (linear) density threshold for collapse in the spherical collapse model:
\begin{equation}
    \nu = \frac{\delta_{\rm c}^2}{D^2(z)\sigma^2}
\end{equation}
where $\delta_{\rm c}=1.686$ and $D(z)$ is the growth factor in cosmological linear perturbation theory:
\begin{equation}
D(z) = \frac{H(z) \int_0^{1/(1+z)} \frac{da}{a^3H^3(a)}}{H_0 \int_0^1 \frac{da}{a^3H^3(a)}}    
\end{equation}
where $H$ is the Hubble parameter.

We note that we need to introduce a correction to the formalism described above since our simulation suite uses a zoom-in technique with a high-resolution volume that is in fact over-dense relative to the mean cosmic volume. Notice that this bias in the mean overdensity over the simulated volume is present even after using the technique described in \citet{Bohr2020} in which the high-resolution region within the larger parent cosmological box is chosen to match as closely as possible the power spectrum of the (lower resolution) parent box in the overlaping scales (see Fig. 2 of \citealt{Bohr2020}).

Due to this bias, the mass function given by Eq. \eqref{eq:hmf} is not directly comparable to the halo mass function extracted from our simulations. It needs to be adjusted for finite volume effects in two ways \citep[see also][]{Sheth2002}: (i) the mass variance has to be corrected for the mass variance of the high-resolution subregion of mass $M_{\rm sub}$
\begin{equation}
    \sigma^2(M) \rightarrow \sigma^2(M) - \sigma^2(M_{\rm sub}),
    \label{sigma_cor}
\end{equation}
and (ii) the threshold for collapse needs to be shifted by the overdensity of the subregion $\delta_{\rm sub}$:
\begin{equation}
    \delta_c \rightarrow \delta_c - \delta_{\rm sub}.
    \label{delta_cor}
\end{equation}

For the window function $\widetilde{W}_R(k)$ in Eq.~\eqref{eq:sigma}, the top-hat filter is the common and successful choice when studying CDM, while a sharp-$k$ filter gives better results for WDM \citep[]{Schneider2013}, but neither seems to accurately account for DAO features in the linear power spectrum \citep{Schewtschenko2015}. \citet{Leo2018smooth} proposed a smooth-$k$ space filter, which does not abruptly cut off like the sharp-$k$ filter, but transitions more smoothly according to:
\begin{equation}
    \widetilde{W}^{\rm smooth}_R(k) = \frac{1}{1+\left(\frac{kR}{c_{\rm W}}\right)^\beta}, \label{eq:smooth}
\end{equation}
where the two free parameters $\beta$ and $c_{\rm W}$ control how sharp the cut off transition is and re-scale the size of the collapsing region ($\widetilde{R}=R/c_W$), respectively. \citet{Sameie2019} used this filter to study the halo mass function of weak DAO models from previous ETHOS simulations ($h_{\rm peak}=0.2$; based on \citealt{Vogelsberger2016}) and found a relatively good agreement. In this work, we use this smooth filter to study the suite of ETHOS simulations from \citet{Bohr2020} within the ($h_{\rm peak}$,$k_{\rm peak}$) parameter space. 

\subsection{EPS formalism applied to ETHOS models}
\begin{figure}
    \centering
    \includegraphics[width=\columnwidth]{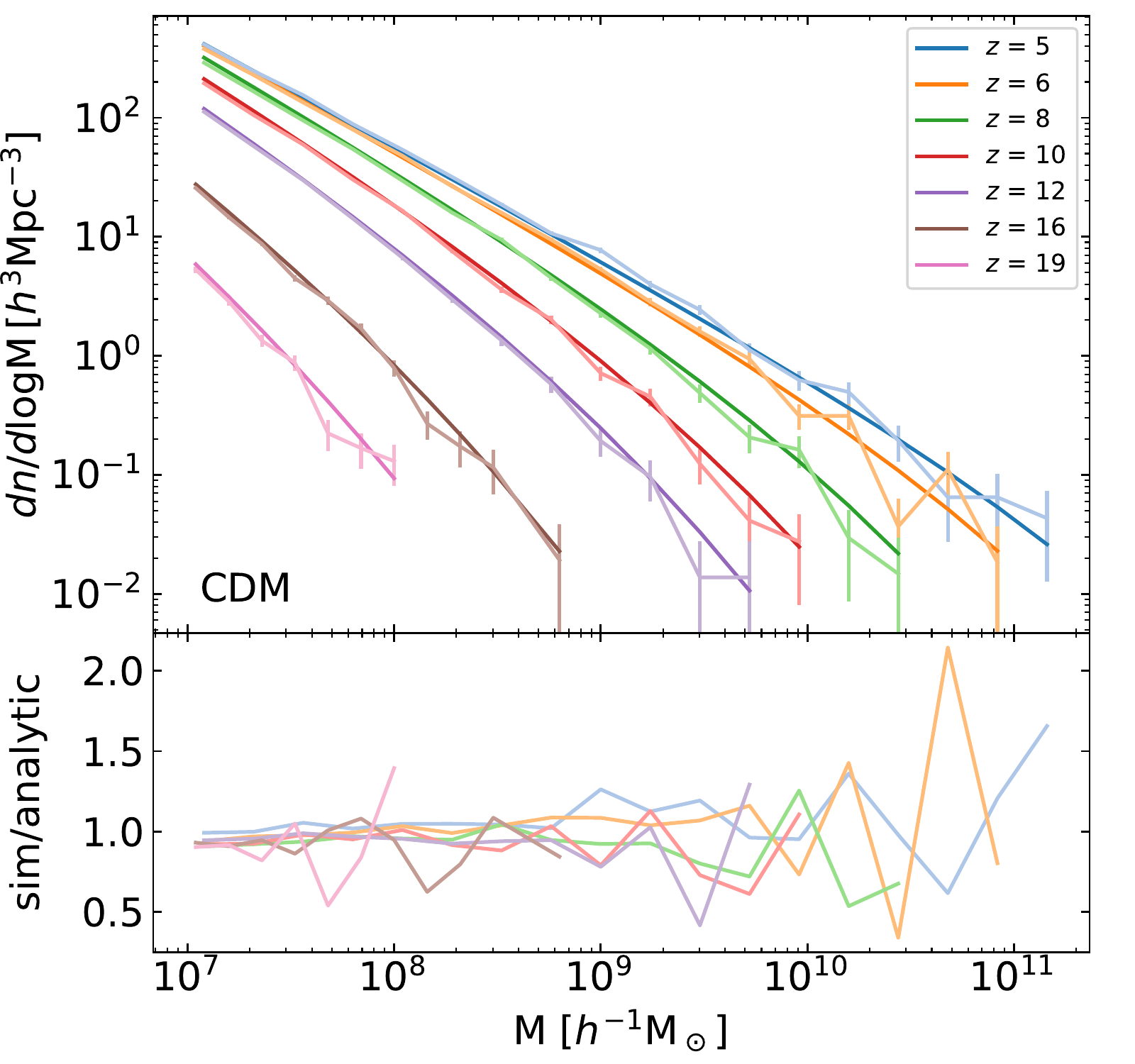}
    \caption{Halo mass function for the CDM model at different redshifts ($z\geq5$) according to the colours in the legend. The light coloured lines with error bars are measurements from the CDM simulation in \citet{Bohr2020}; the error bars are Poissonian. The 
    dark coloured lines without error bars are computed from the EPS halo mass function (Eq.~\ref{eq:hmf}) corrected for finite volume effects (Eqs.~\ref{sigma_cor}$-$\ref{delta_cor}) and with smooth-$k$ space window function (Eq.~\ref{eq:smooth}). The best-fit parameters of the EPS halo mass function are 
    $A=0.3658$, $\beta=3.46$, $c_{\rm W}=3.79$. The bottom panel shows the ratio between the simulation and the EPS results at each redshift.}
    \label{fig:hmf_cdm}
\end{figure}

\begin{figure}
    \centering
    \includegraphics[width=\columnwidth]{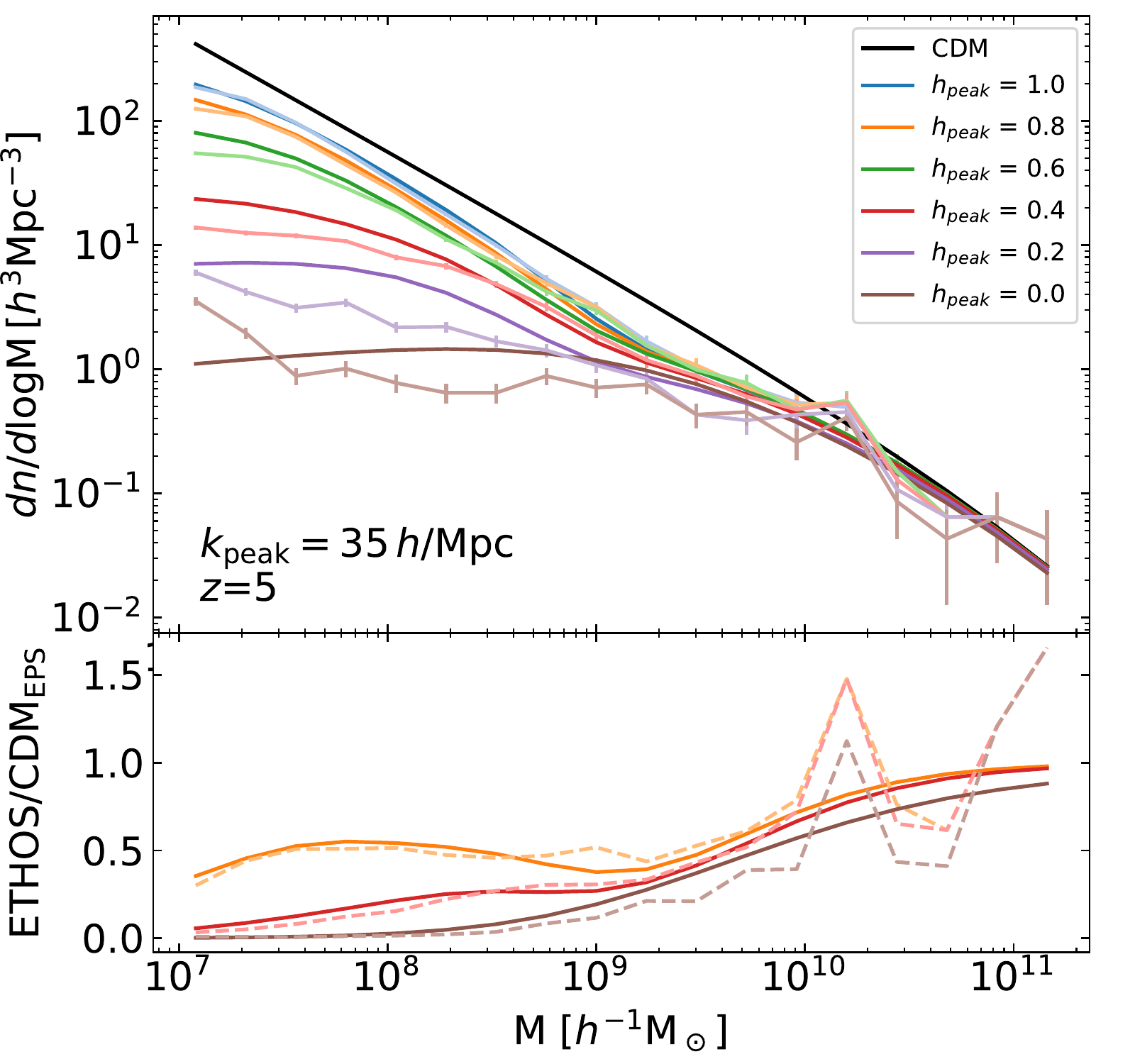}
    \caption{Halo mass function at $z=5$ for the ETHOS simulations of \citet{Bohr2020} having the same $k_{\rm peak}=35$\hmpc (i.e. the same linear power spectrum cutoff scale), but with different values of $h_{\rm peak}$ according to the different colours in the legend, from WDM ($h_{\rm peak}=0$), to models with strong DAOs ($h_{\rm peak}=1$). The light coloured lines with error bars are the simulation results, while the dark coloured lines are produced with the EPS model as described in the text and the caption of Fig.~\ref{fig:hmf_cdm}. For the lowest $h_{\rm peak}$, the onset of spurious haloes \citep[see][]{Wang2007} is visible just above $10^7$\msunh. The bottom panel shows the ratio of ETHOS with respect to the CDM EPS result for a selection of models as given by the colours, with the case of the simulations (EPS models) shown with dashed (solid) lines.}
    \label{fig:hmf_hpeaks}
\end{figure}

\begin{figure}
    \centering
    \includegraphics[width=\columnwidth]{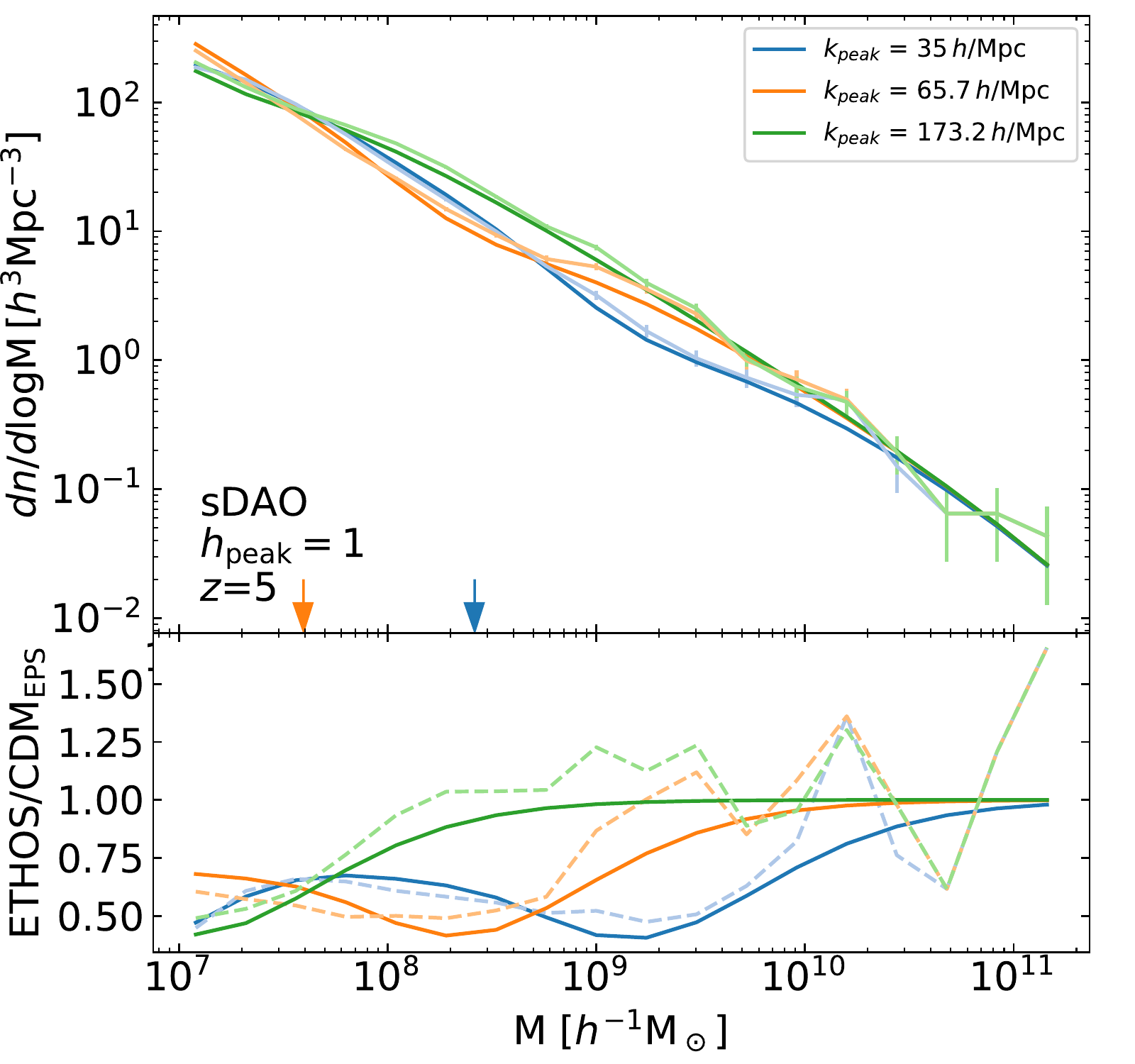}
    \includegraphics[width=\columnwidth]{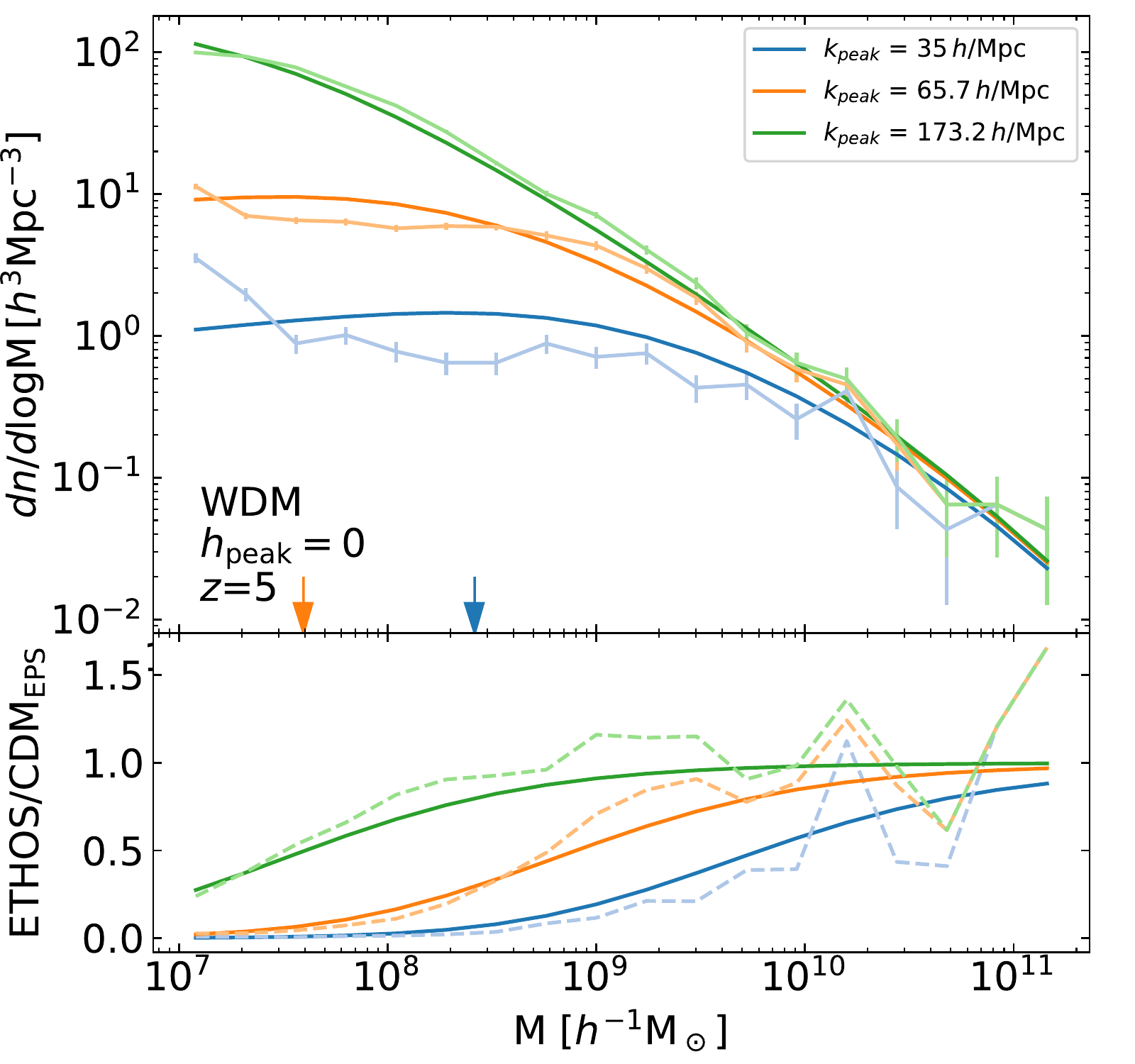}
    \caption{Halo mass function at $z=5$ for the ETHOS simulations of \citet{Bohr2020} having the same DAO amplitude ($h_{\rm peak}=1$, sDAO on the top, and $h_{\rm peak}=0$, WDM on the bottom), but with different values of $k_{\rm peak}$ (the corresponding mass is indicated with arrows) according to the different colours in the legend. The light coloured lines with error bars are the simulation results, while the dark coloured lines are produced with the EPS model as described in the text and the caption of Fig.~\ref{fig:hmf_cdm}. For the WDM models, the onset of spurious haloes \citep[see][]{Wang2007} is visible just above $10^7$\msunh. The respective bottom panels show the ratio of ETHOS with respect to the CDM EPS result, with the case of the simulations (EPS models) shown with dashed (solid) lines.}
    \label{fig:hmf_kpeaks}
\end{figure}

With all the previous considerations, we fit the free parameters in the EPS mass function simultaneously to all our ETHOS simulations
in the range $h_{\rm peak}=0-1$ and $k_{\rm peak}=35-300$\hmpc (plus CDM), and across the redshift range $z=5-12$ by minimizing the $\chi^2$. We fit the halo mass function for $M>10^7$\msunh or $M>M_{\rm lim}$ for models with $M_{\rm lim}>10^7$\msunh, where $M_{\rm lim}$ is the limiting mass for spurious haloes as defined in \citet{Wang2007}. For the mass of our simulated haloes, we use $M_{200}=4\pi /3 r_{200}^3 200\rho_{\rm c}$, where the virial radius $r_{200}$ is defined as the radius at which the enclosed density is 200 times the critical density of the Universe $\rho_{\rm c}$. We find the best-fitting parameters to be: $A=0.3658$, $\beta=3.46$, $c_{\rm W}=3.79$. The agreement between the best-fit parameters of the EPS mass function and the simulation results can be seen in
Figs.~\ref{fig:hmf_cdm}$-$\ref{fig:hmf_z}, where the faded lines with error bars are the result from the simulations and the solid lines are the analytic predictions using the same best-fit parameters as in the CDM case (given in the caption of Fig.~\ref{fig:hmf_cdm}).


Fig.~\ref{fig:hmf_cdm} shows the CDM halo mass function and it is clear that our EPS implementation results is in an overall good fit to the simulation data across a wide range of redshifts ($5\leq z\leq19$). The scatter at the largest halo masses at a given redshift in the simulation results is expected and comes from low-number statistics, given the relatively small volume of our zoom-in simulations. In the mass range where the sampling error is small, the typical mismatch between the EPS modelling and the simulations is $\lesssim10\%$.

Fig.~\ref{fig:hmf_hpeaks} shows the models with the smallest value of $k_{\rm peak}=35$\hmpc in our simulations for the full range of $h_{\rm peak}=0-1$. These models correspond to linear power spectra with the largest cutoff-scale. By looking at the upper panel, it is clear that for sDAO models ($h_{\rm peak}\gtrsim0.6$) the analytic prediction can accurately reconstruct the halo mass function across all masses. The small-scale suppression in the linear power spectrum relative to CDM results in a deficit in the abundance of low-mass haloes, which is captured quite well by the EPS formalism, both in the cutoff mass-scale, and even in the subsequent oscillations observed at smaller masses. For the wDAO ($h_{\rm peak}\lesssim0.6$) and WDM ($h_{\rm peak}=0$) models on the other hand, only the general cut-off is captured by the analytic prediction, while the amplitude and details at small masses are slightly over-predicted and not captured as well. The bottom panel of Fig.~\ref{fig:hmf_hpeaks} shows the ratio between the halo mass function of the ETHOS model (simulation in faded lines, and EPS predictions in dark coloured lines) to that of the CDM EPS prediction. If we just compare a given ETHOS model to CDM, 
the suppression of low mass haloes for WDM and wDAO models far outweighs the differences between the simulation result and the EPS formalism. We note that overall, our results in regards to the mismatch between the EPS formalism and the case $h_{\rm peak}=0.2$ (belonging to the class of ETHOS models studied in \citealt{Vogelsberger2016}; see \citealt{Bohr2020}) is in general agreement with the high redshift results of \citet[][]{Sameie2019} who directly studied the ETHOS simulations of \citet{Vogelsberger2016}. 

We also notice that the models with $h_{\rm peak}\leq0.2$ suffer from the presence of spurious haloes due to discreetness effects; a well known artifact in models where the linear power spectrum is well below the unavoidable Poisson noise present in the creation of the initial conditions \citep[see][]{Wang2007}. For these models, the halo mass function starts rising artificially towards the smallest masses just below a few times $10^7$\msunh. We notice that the mass scale where spurious haloes becomes apparent in the halo mass function of our simulations is roughly in agreement with the limiting mass formula for discreteness effects given by \citet{Wang2007}\footnote{$M_{\rm lim} = 10.1 \times \bar{\rho} d k_{\rm p}^{-2}$, where $\bar{\rho}$ is the mean density of the Universe, $d$ is the mean interparticle separation, and $k_{\rm p}$ is the wavenumber at which the initial dimensionless power spectrum reaches its maximum.}. For instance, for our most extreme WDM model, the limiting mass according to \citet{Wang2007} is $1.6\times 10^8$\msunh, whereas we see a clear artificial increase in the halo mass function at about half this value. For most of the models we analyse, the limiting mass is significantly lower than that of the extreme WDM model, and since the range of masses we are interested on is above this maximum limiting value, we will not discuss the presence of spurious haloes any further.
Notice that in Section~\ref{sec:inner} below we only analyse the inner structure of haloes having a mass at least an order of magnitude larger than the mass where spurious haloes starts to become apparent in the halo mass function.

The behaviour of the halo mass function for a fixed $h_{\rm peak}$ but with different $k_{\rm peak}$ values (i.e. effectively different cutoff scales in the linear power spectrum) is shown in Fig.~\ref{fig:hmf_kpeaks}. The top panel exemplifies the sDAO models ($h_{\rm peak}=1$) while WDM models ($h_{\rm peak}=0$) are shown in the bottom panel. The EPS formalism remarkably captures the shift of the cut-off mass for different values of $k_{\rm peak}$; the signature of the DAOs in the halo mass function is also well reproduced by the model. In light of this agreement with the EPS formalism, we can say that this results confirms the expectation that the halo mass at which the cut-off occurs is directly connected to the mass within a radius proportional to the DAO scale $k_{\rm peak}$. On the other hand, for the WDM models (bottom panel of Fig.~\ref{fig:hmf_kpeaks}) it is especially noticeable that the agreement between the EPS model and the simulation becomes progressively better with increasing $k_{\rm peak}$. That behaviour is expected as the models approach CDM with increasing $k_{\rm peak}$.

Finally, Fig.~\ref{fig:hmf_z} shows that also the redshift evolution of sDAO models is well captured by the EPS formalism in a way that is essentially as good as for CDM (see Fig.~\ref{fig:hmf_cdm}). In the ratio relative to CDM (bottom panel of Fig.~\ref{fig:hmf_z}), shown only for three redshifts, it is also visible that the deficit of low-mass haloes (relative to CDM) is higher at larger redshifts and progressively decreases towards lower redshifts. 

Overall, we conclude that the halo mass function predicted by the EPS formalism, corrected by finite volume effects and with the smooth-$k$ space filter works very well for sDAO models ($h_{\rm peak}=0.6-1$) across all probed masses. The formalism however, over-predicts the small mass abundance for WDM and wDAO models although the difference with respect to CDM is still reasonably captured. Finally, we note that we were also able to reconstruct the halo mass function of the wDAO ETHOS models presented in \citet{Vogelsberger2016} in the redshift range studied here and found a reasonable agreement with our EPS modelling, in line with what was described above for wDAO models.

\subsection{Shape of the Halo Mass Function for ETHOS models}
\begin{figure}
    \centering
    \includegraphics[width=\columnwidth]{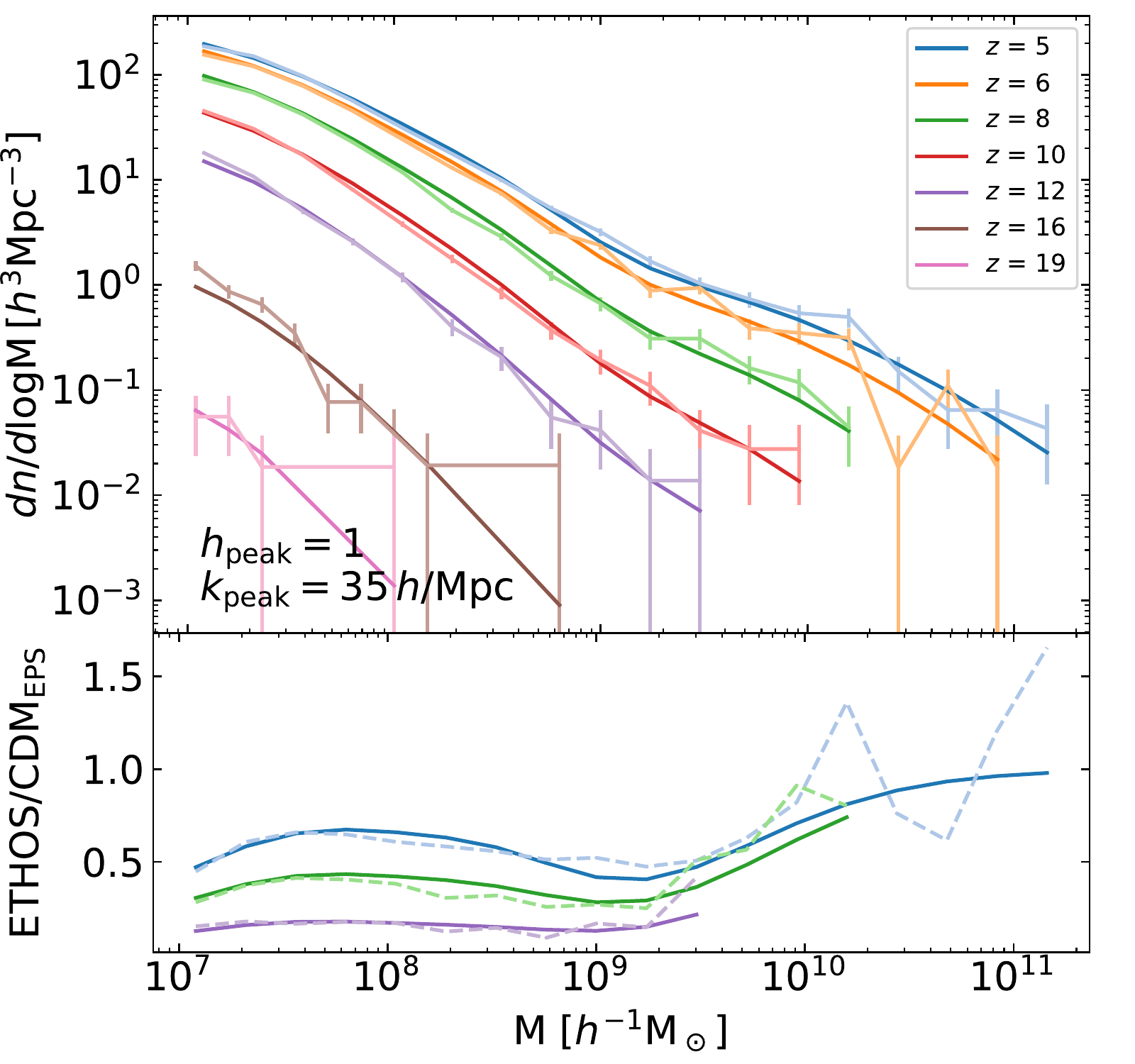}
    \caption{Halo mass function for a sDAO model ($h_{\rm peak}=1$, $k_{\rm peak}=35$\hmpc) at different redshifts ($z\geq5$) according to the different colours in the legend. The light coloured lines with error bars are the simulation results, while the dark coloured lines are produced with the EPS model as described in the text and the caption of Fig.~\ref{fig:hmf_cdm}. The bottom panel shows the ratio of the sDAO model with respect to the CDM EPS result for a selection of redshifts as given by the colours, with the case of the simulations (EPS models) shown with dashed (solid) lines.}
    \label{fig:hmf_z}
\end{figure}

Figures \ref{fig:hmf_hpeaks}$-$\ref{fig:hmf_z} make clear that the shape of the halo mass function for models displaying DAOs in their linear matter power spectrum differs significantly from either the WDM or CDM case. While WDM ($h_{\rm peak} = 0$) mass functions are characterized by a uniform and monotonic suppression below a given mass scale (usually parameterized by their half-mode mass), DAO models display non-monotonic mass functions for which the initial (higher mass) suppression is followed by a localized feature where the mass functions converge back towards the CDM amplitude before decaying again on even smaller mass scales. This localized feature is clearly visible in the lower panel of Fig.~\ref{fig:hmf_hpeaks} where we see that it becomes more prominent as $h_{\rm peak}$ increases. The presence of this feature is a direct consequence of the early-universe acoustic waves propagating in the dark sector for these ETHOS models, which later become imprinted in the dark matter density field once the latter decouples from the radiation bath. These frozen density waves then provide a slight enhancement of the dark matter fluctuations field once smoothed over a scale corresponding roughly to the DAO scale, hence leading to an excess of halos as compared to a WDM model with a similar initial suppression. 

The upper panel of Fig.~\ref{fig:hmf_hpeaks} makes clear that the presence of the DAO feature can change the halo mass function by orders of magnitude compared to the simpler WDM case. Indeed, while the sDAO model with $h_{\rm peak}=1$ and the WDM ($h_{\rm peak}=0$) model deviate from the CDM case in a very similar fashion near ${\rm M}=10^{10}h^{-1}{\rm M}_\odot$, the sDAO model then reconverges towards the CDM mass function, resulting in an abundance of $10^7 h^{-1}{\rm M}_\odot$ haloes that is more than 2 orders of magnitude greater than that of WDM. The peculiar shape of the ETHOS halo mass function means that constraints on dark matter physics based on the abundance of small-scale structure (using, e.g.~, lensing or satellites) cannot straightforwardly be applied to these models.

\section{The inner structure of ETHOS haloes at high redshift ($z=5$)} \label{sec:inner}

Having described and analysed the abundance of haloes at high-z within ETHOS in the context of the EPS formalism, we now look at the dark matter distribution within these haloes. In particular, we study the spherically-averaged density profile of ETHOS haloes at high-z and focus on the concentration-mass relation in the context of the halo assembly model of \citet{Ludlow2016}. 

\subsection{Density profile}

The near-universality of CDM haloes has been well established since the seminal papers of \citet{NFW1996,NFW1997}. The well-known 2-parameter Navarro-Frenk-White (NFW) profile has been shown to be a remarkably well fit to the spherically-averaged radial density profile of CDM haloes. Although more recent, higher resolution simulations show that other profiles such as the Einasto profile provide an even better fit to the structure of simulated haloes (e.g. \citealt{Springel2008}), the simplicity and accuracy of the NFW profile remains valid. This is particularly relevant when one considers that the NFW profile effectively becomes a function of one free parameter since there is a tight correlation (monotonically decreasing) between the virial mass of the halo and its concentration \citep[e.g.][]{Bullock2001,Eke2001,Neto2007,Prada2012,Ludlow2014,SC2014,Klypin2016,Uchuu2020,Wang2020}. The concentration parameter for NFW haloes is defined as $c=r_{200}/r_{\rm s}$, where $r_{200}$ is the virial radius, 
and $r_{\rm s}$ is the scale radius, which for the NFW profile coincides with $r_{-2}$, the radius where the logarithmic slope of the density profile is $-2$. 

\begin{figure}
    \centering
    \includegraphics[width=\columnwidth]{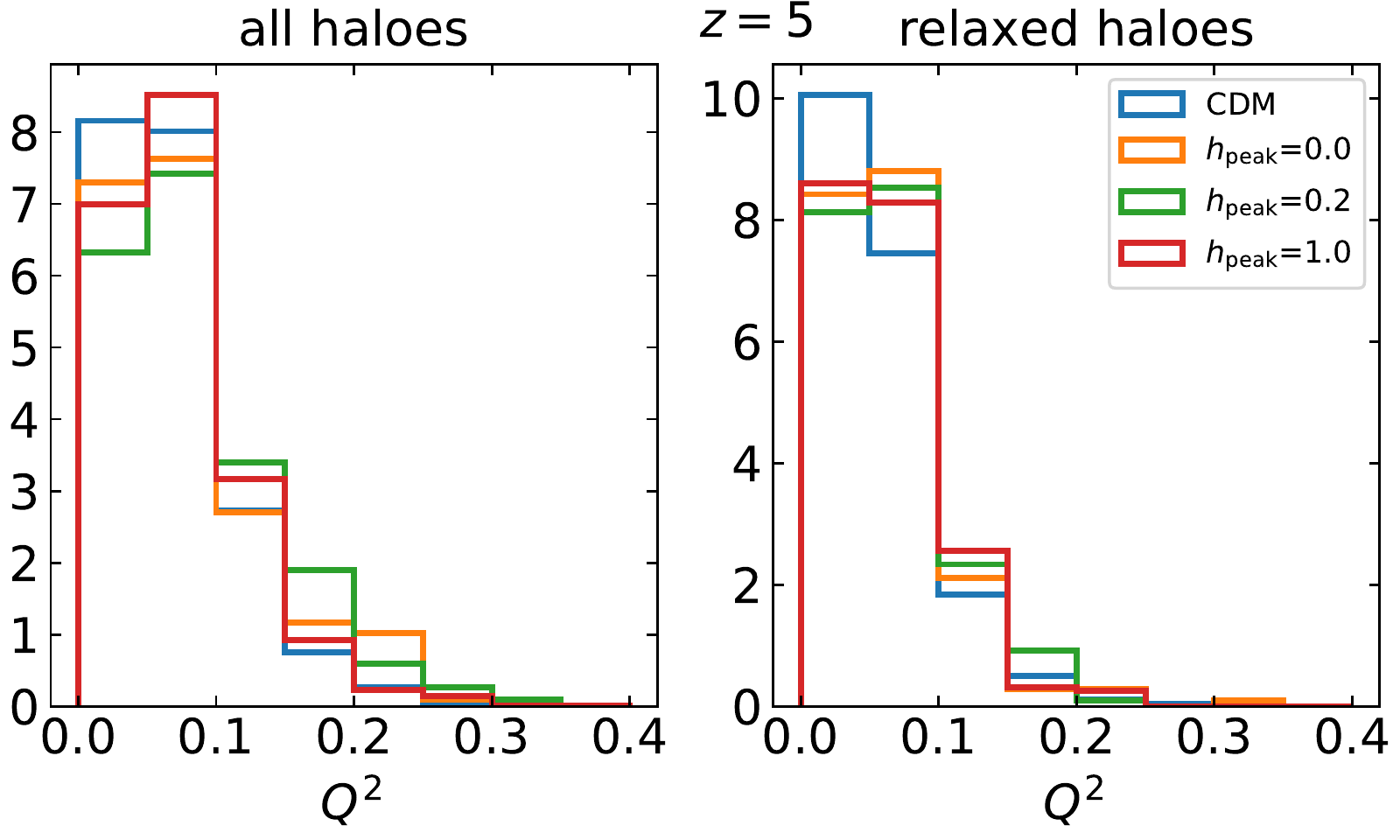}
    \caption{Goodness of the fit of NFW profiles for CDM, WDM ($h_{\rm peak}=0$), wDAO and sDAO models (all three non-CDM models with $k_{\rm peak}=35$\hmpc) measured by $Q^2$ (Eq.~\ref{eq:Q2}). The left panel shows all simulated haloes with more than 5000 particles, while the right panel considers in addition only relaxed haloes (according to the criteria in Appendix~\ref{sec:relaxation}).}
    \label{fig:Q2}
\end{figure}

The near-universality of the NFW profile extends to the WDM case as well, where it has been shown that WDM haloes are also well described by this profile, albeit with lower concentration than the CDM counterpart at fixed mass \citep[e.g.][]{Lovell2014,Bose2016}. We thus begin this section by analysing if the NFW profile provides a good fit to ETHOS haloes in general, that is, we explore if the near-universality of this profile extends as well to models with DAOs.
To quantify this we create density profiles for all haloes with at least 5000 particles\footnote{This limit is used to obtain a robust sampling of the spatial structure of a halo.} in a given ETHOS model using concentric shells from the centre of each halo, defined from the minimum of the halo potential. The shells are binned
logarithmically in the range $6\epsilon/r_{200} < r/r_{200} < 3$, where six times the smoothing length $\epsilon$ is the convergence limit of our simulations (see Appendix~\ref{sec:convergence}). 
We then fit the simulated profiles with the NFW profile by minimizing the following quantity:
\begin{equation}
    Q^2 = \frac{1}{N_{\rm bins}} \sum_{i=1}^{N_{\rm bins}} \left(\ln\rho_i - \ln\rho_i^{\rm NFW}\right)^2, \label{eq:Q2}
\end{equation} 
where $N_{\rm bins}=50$ \citep[see][]{Navarro2010}. 
The left panel of Figure~\ref{fig:Q2} shows the $Q^2$ distribution for all haloes with more than 5000 particles for the CDM, WDM, wDAO and sDAO models. While a slightly higher fraction of CDM haloes are in the smallest $Q^2$ bin, the shape and width
of the distribution is quite similar across all DM models.
If we only look at relaxed haloes (right panel of Fig.~\ref{fig:Q2}; see Appendix~\ref{sec:relaxation} for the relaxation criteria we used), we have a narrower distribution with an even higher fraction of haloes with small $Q^2$, which are therefore well described by a NFW profile.  

\subsection{Concentration-mass relation}

\citet{Ludlow2016} developed an analytic model for the concentration-mass relation based on EPS theory and applied it to CDM and WDM models (this model is an extension of the one developed earlier in \citealt{Ludlow2014}). Their model assumes that the mean inner density within the scale radius $\left<\rho_{-2}\right>$ is proportional to the critical density of the Universe at an assembly redshift $z_{-2}$:
\begin{equation}
    C \left(\frac{H(z_{-2})}{H(z_0)}\right)^2 = \frac{\left<\rho_{-2}\right>}{\rho_0} = 200 c^3 \frac{\ln(2)-0.5}{\ln(1+c)-c/(1+c)} ,
    \label{eq:concentration0}
\end{equation}
where the second equality is only valid for NFW profiles, $C$ is a free parameter.
Secondly, the model assumes that the assembly redshift is defined as the redshift when the enclosed mass within the scale radius $M_{-2}$ of the descendant halo was first assembled into progenitors having a mass larger than a fraction $f$ of the descendant,
and is given by
\begin{equation}
    \operatorname{erfc} \left(\frac{\delta_c(z_{-2})-\delta_c(z_0)}{\sqrt{2(\sigma^2(f\times M) - \sigma^2(M)}}\right) = \frac{M_{-2}}{M_0} = \frac{\ln(2)-0.5}{\ln(1+c)-c/(1+c)} , \label{eq:concentration}
\end{equation}
where the second equality is only valid for NFW haloes and $\delta_{\rm c}(z)=\delta_{\rm c}/D(z)$ is the redshift dependent critical density for collapse. The left hand side of Eq.~\ref{eq:concentration} corresponds to the collapsed mass fraction in EPS theory \citep{Lacey1993}. Across the paper, we use $C=575$ and $f=0.02$ for the free parameters in Eqs.~\eqref{eq:concentration0}$-$\eqref{eq:concentration}.

\begin{figure}
    \centering
    \includegraphics[width=\columnwidth]{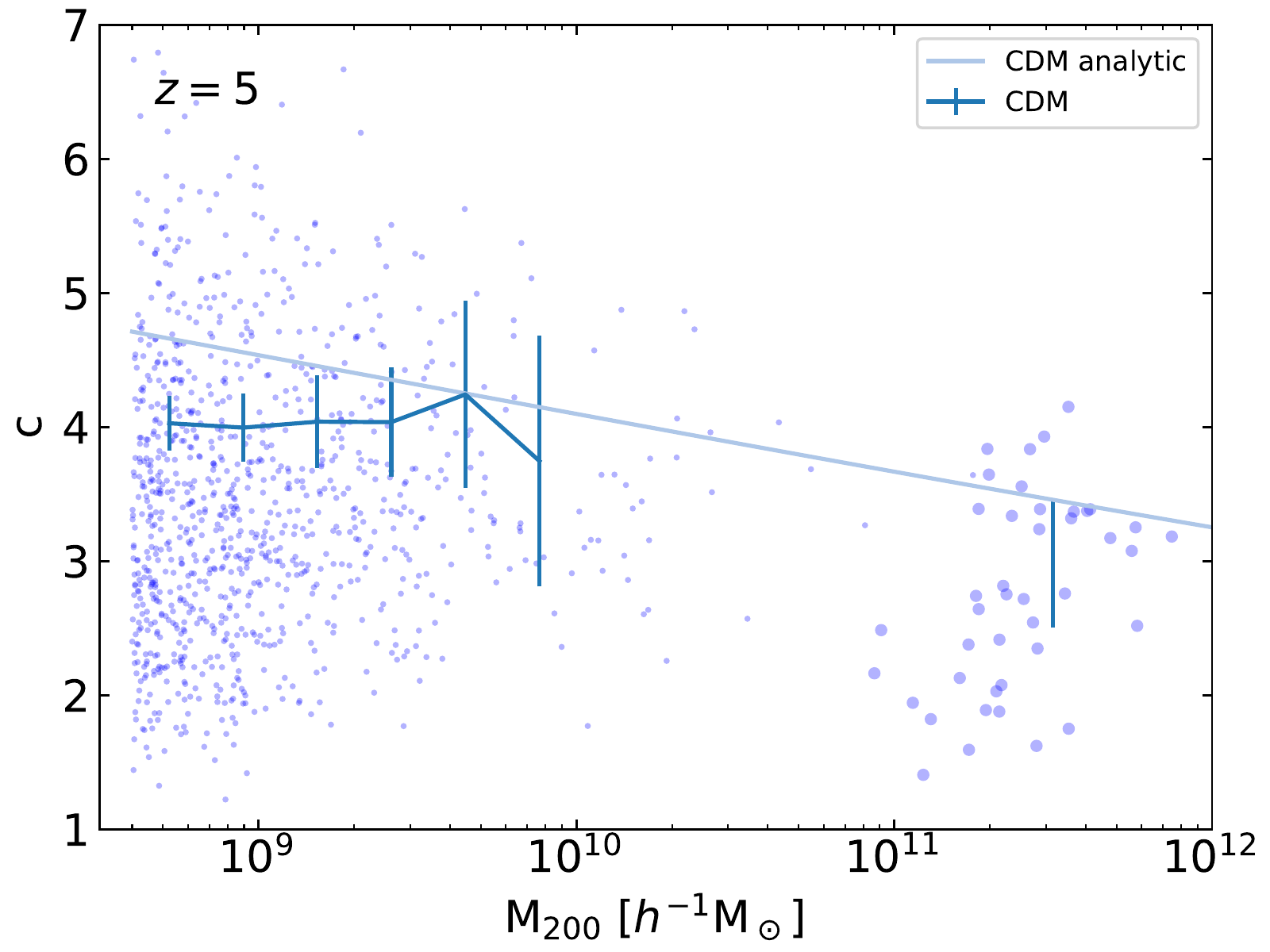}
    \caption{Concentration mass relation for CDM haloes at $z=5$. The scatter points represent individual measurements of the NFW concentration for all relaxed CDM haloes while the lines with Poisson error bars correspond to the median concentration at that mass. The median is computed at four equally sized bins for $M=10^9-10^{10}\,M_\odot/h$ from high-res haloes and one bin for $M=10^{11}-10^{12}\,M_\odot/h$ from low-res haloes. The lines correspond to the concentration computed using the analytic model of \citealt{Ludlow2016} with a smooth-$k$ space filter (see Eq.~\eqref{eq:concentration}).}
    \label{fig:concentration_cdm}
\end{figure}

\begin{figure}
    \centering
    \includegraphics[width=\columnwidth]{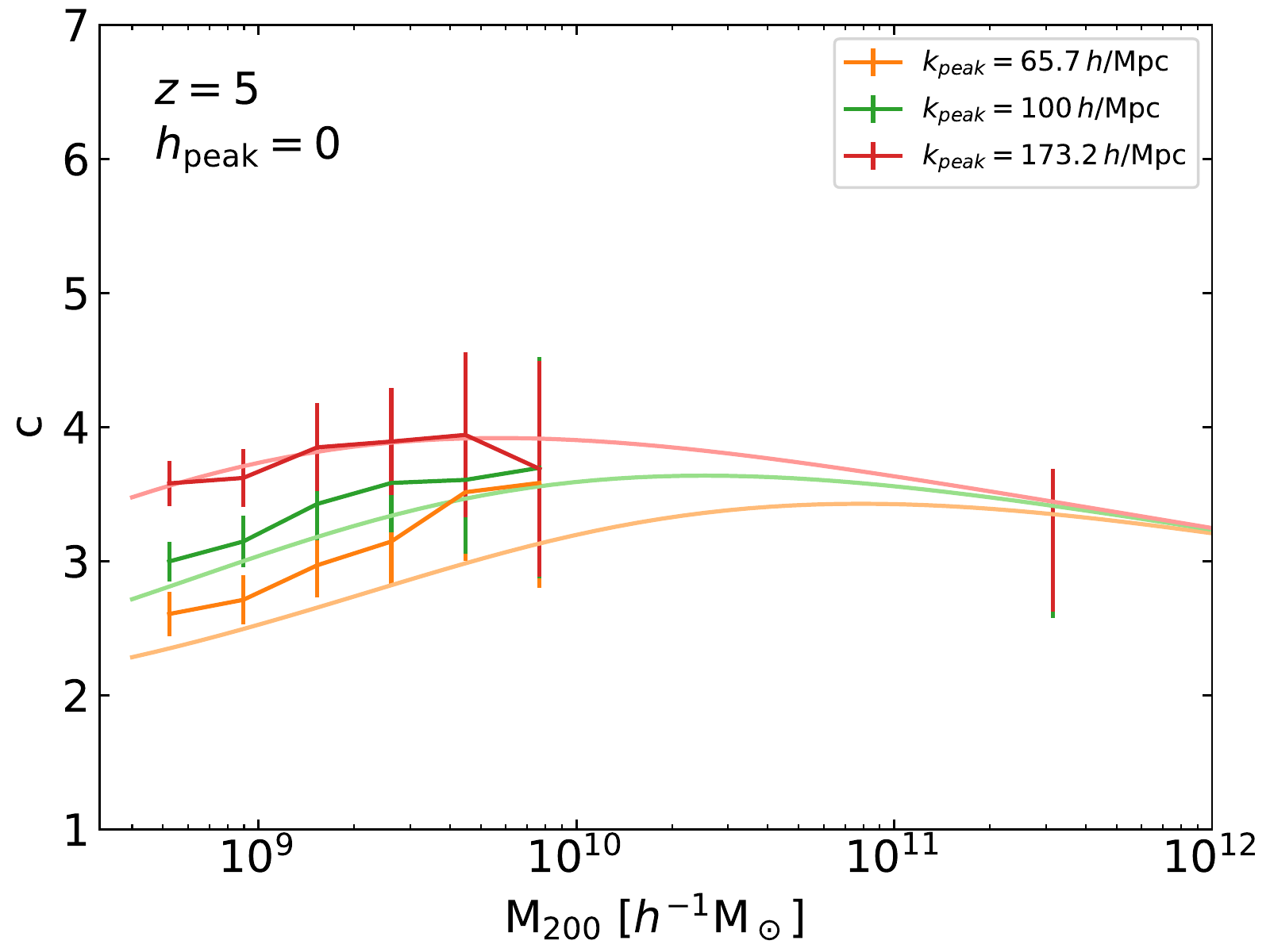}
    \caption{Concentration mass relation for WDM haloes at $z=5$. The lines with Poisson error bars correspond to the median concentration of haloes at that mass. The median is computed at four equally sized bins for $M=10^9-10^{10}\,M_\odot/h$ from high-res haloes and one bin for $M=10^{11}-10^{12}\,M_\odot/h$ from low-res haloes. The lines correspond to the concentration computed using the analytic model of \citealt{Ludlow2016} with a smooth-$k$ space filter (see Eq.~\ref{eq:concentration}).}
    \label{fig:concentration_wdm}
\end{figure}

\begin{figure}
    \centering
    \includegraphics[width=\columnwidth]{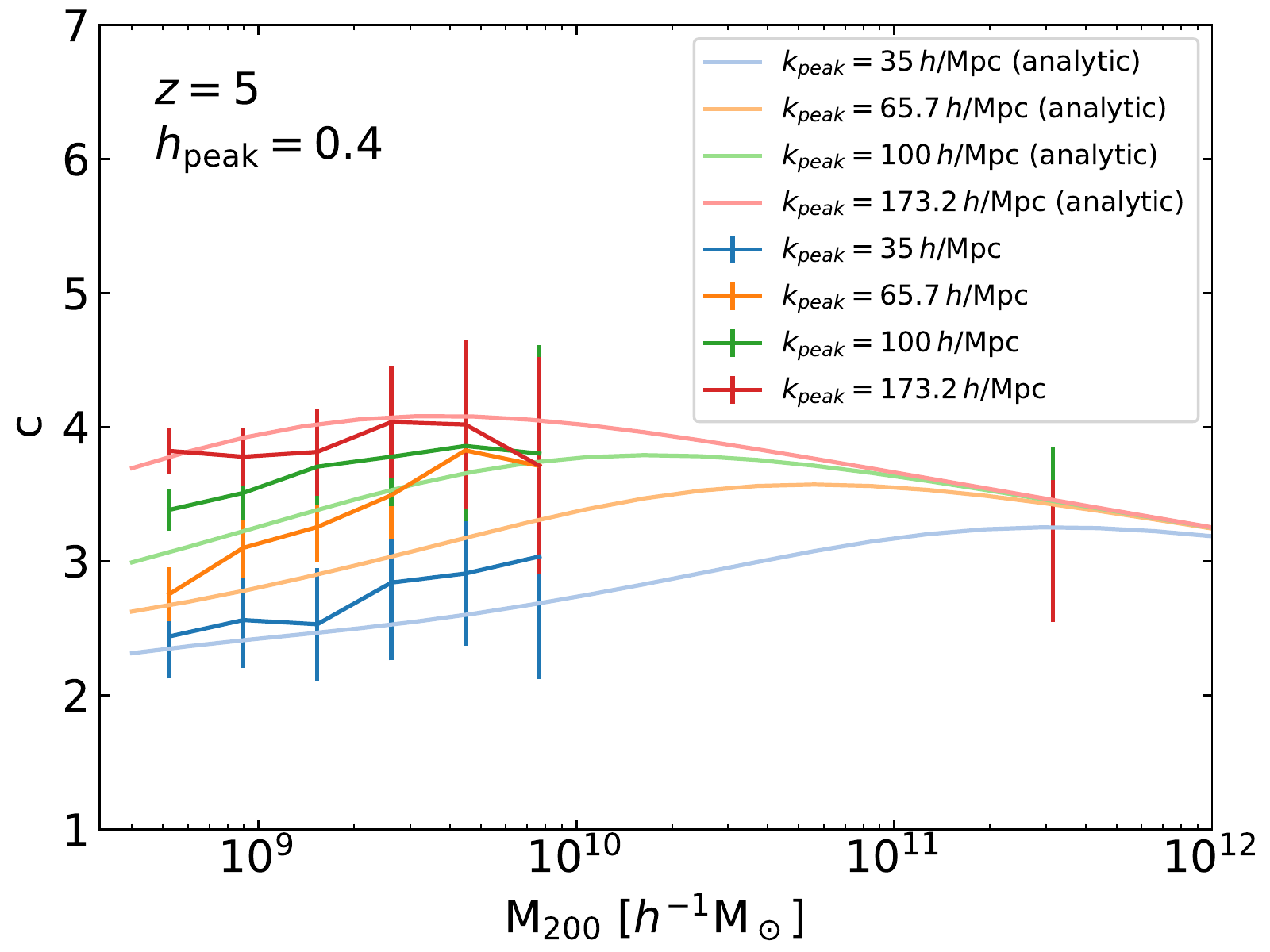}
    \includegraphics[width=\columnwidth]{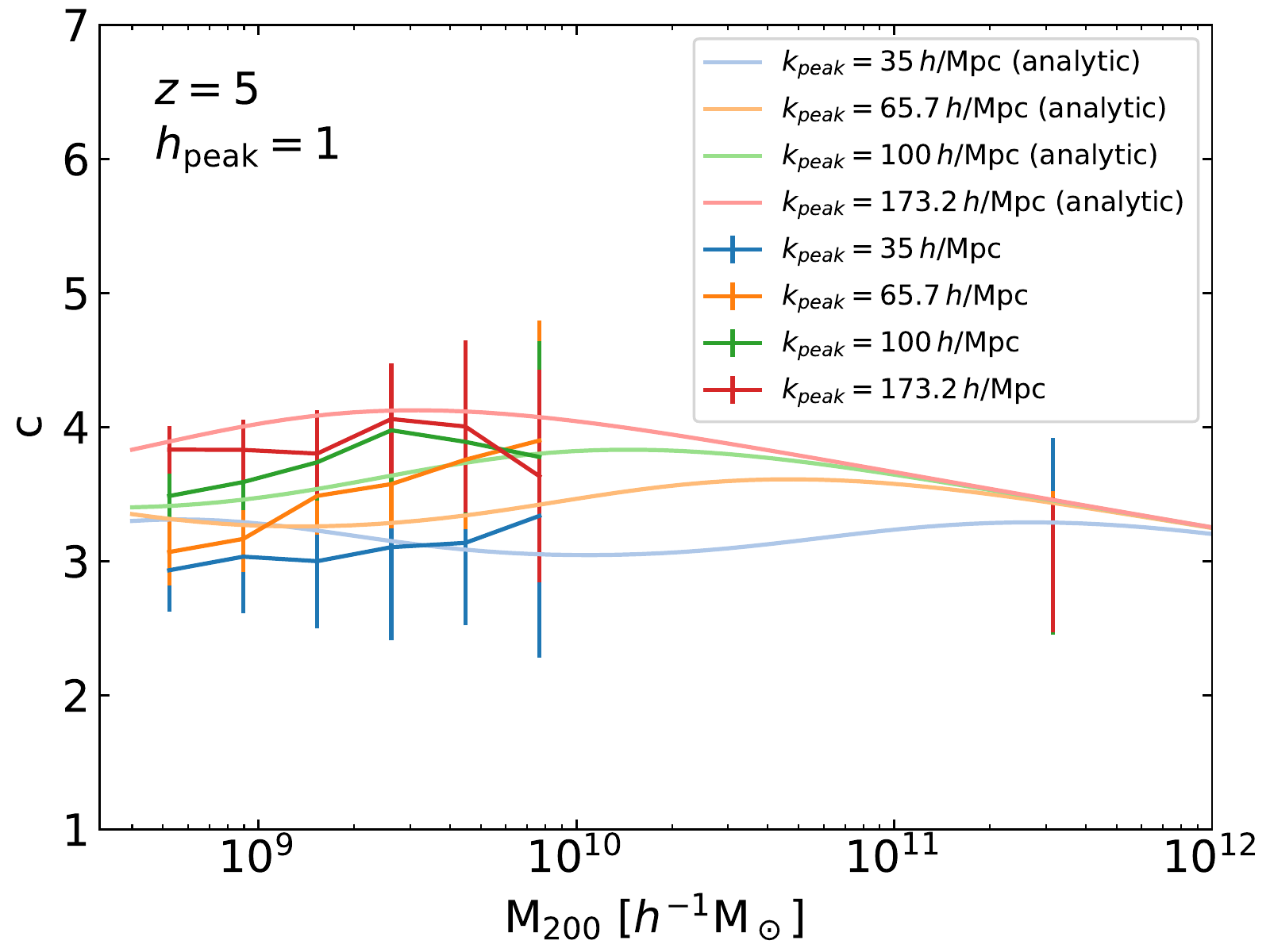}
    \caption{Concentration mass relation for wDAO (top) and sDAO (bottom) haloes at $z=5$. The lines with Poisson error bars correspond to the median concentration of haloes at that mass bin. The median is computed at four equally sized bins for $M=10^9-10^{10}\,M_\odot/h$ from high-res haloes and one bin for $M=10^{11}-10^{12}\,M_\odot/h$ from low-res haloes. 
    The lines correspond to the concentration computed using the analytic model of \citealt{Ludlow2016} with a smooth-$k$ space filter (see Eq.~\ref{eq:concentration}).}
    \label{fig:concentration_dao}
\end{figure}

Figures~\ref{fig:concentration_cdm}$-$\ref{fig:concentration_dao} show the concentration-mass relation at $z=5$ for CDM, WDM, wDAO and sDAO models obtained from our simulations and the analytic model described above using a smooth-$k$ filter for all models. For these plots, we only took relaxed haloes into account and binned the haloes in the high-resolution region (hereafter high-res haloes) into four bins, equally sized in logarithmic mass bins 
in the range $10^9-10^{10}$\msunh. We obtained the concentration for each bin by taking the median density profile of all relaxed haloes within that bin (stacking the profiles by re-scaling the radius to the virial radius and only up to $r/r_{200}=0.8$) and fitting a NFW profile to it. The irregularities of individual haloes are smoothed out in this way. Additionally, we have used data from the low-resolution regions of the simulations (hereafter low-res haloes), combining the low-res haloes with ${\rm M}=10^{11}-10^{12}$\msunh into one mass bin to serve as high-mass anchor point, when comparing the analytic model to our simulations.

For CDM, we use the smooth-$k$ filter when calculating the mass variance $\sigma^2$ in Eq.~\eqref{eq:concentration} and
Fig.~\ref{fig:concentration_cdm} shows that, as expected, the analytic model and our simulations are in reasonable agreement within the Poisson sampling errors. However, at low masses we observe that the concentration remains flat in the simulations, while the model predicts a monotonically decreasing concentration.

For WDM (Fig.~\ref{fig:concentration_wdm}), we also use the smooth-$k$ filter and the simulations are captured well in this way. The low-mass behaviour is followed closely by the analytic model and the high-mass anchor point is also in good agreement. 
Even though the analytic model and the simulation results agree within the Poisson noise, we notice a trend that the model under-predicts the concentration for small $k_{\rm peak}$ and over-predicts for large $k_{\rm peak}$. This suggests that the scaling with $k_{\rm peak}$ is not captured completely accurately by the model. However, the number of haloes in our simulations is not large enough to 
fully trust this trend. We note that the WDM model with $k_{\rm peak}=35$\hmpc shows very high concentration values for low mass haloes for which we suspect numerical issues, as this is the model with the most extreme suppression of small scale power. Therefore, we have omitted this model in Fig.~\ref{fig:concentration_wdm}.
We note here, that \citet{Ludlow2016} originally tested their analytic approach for WDM models at lower redshifts than those in our simulations ($z=0-3$). The authors found a good agreement, although their simulation with the smallest WDM particle mass $m_{\rm WDM}=1.5$~keV suggests some inconsistencies at low masses for $z=3$ (green line in Fig.4 of \citealt{Ludlow2016}).

Fig.~\ref{fig:concentration_dao} shows the concentration-mass relation for wDAO (top) and sDAO models (bottom) using the smooth-$k$ space filter (Eq.~\ref{eq:smooth}). As with WDM, the analytic prediction for the DAO models is in reasonable agreement with the simulation. In both cases, we see a similar trend to under-predict the concentration for small $k_{\rm peak}$ and over-predict for large $k_{\rm peak}$. For small $k_{\rm peak}$ however, the concentration of sDAO haloes is predicted to increase again towards the smallest masses, which we do not observe in our simulations. Therefore, it seems that the sDAO features of the most extreme models are not correctly captured by the analytic model of \citet{Ludlow2016}.
We notice however, that across all DAO models, the decrease (soft cutoff) of concentration towards intermediate masses predicted by the analytic model is seen in our simulations, while a predicted increase at lower masses in the sDAO cases is not present in the simulations. We notice that the latter trend (continuous increase of concentraton at low masses), is actually not seen neither in the CDM case (see Fig.~\ref{fig:concentration_cdm}). The over-prediction might therefore not be an explicit problem of the sDAO model.

We can conclude that Figs.~\ref{fig:concentration_wdm} and \ref{fig:concentration_dao} demonstrate that the reduced small-scale power in ETHOS models reduces the concentration of small haloes depending on the value of $k_{\rm peak}$. Furthermore Eq.~\eqref{eq:concentration}, based on the analytic model of \citealt{Ludlow2016} (ultimately based on EPS theory), can capture the concentration-mass relation reasonably well for a wide range of DM models. However, the model struggles to reproduce the small mass behaviour for the most extreme sDAO models and the trend indicates that the scaling with $k_{\rm peak}$ is not captured correctly.
We remark that in order to improve the analytic model, simulations with a larger volume but similar resolution are needed to reduce the sampling errors, while also covering higher mass haloes with
${\rm M}>10^{10}$\msunh.

\section{Conclusions}
\label{sec:conclusion}

Performing dedicated cosmological $N$-body simulations to extract basic but precise measurements of the properties of haloes for 
specific DM models requires access to HPC resources, which can be computationally expensive when a broad exploration of models is desired. Such a broad exploration is essential to cover the range of alternatives to the Cold Dark Matter (CDM) model, which predict a halo population with distinct properties. A relevant category of such alternatives is that of models with a (galactic-scale) primordial cutoff in the linear power spectrum, caused by either the free streaming mechanism (Warm Dark Matter, WDM) or by collisional damping with relativistic species in the early Universe (models with Dark Acoustic Oscillations, DAOs). The difference between these models and CDM increases at low halo masses, which are more affected by the small-scale suppression of power, particularly at high redshift.

For these reasons, in this work we take the simulation suite from \citet{Bohr2020} within the ETHOS framework \citep{Cyr-Racine2016,Vogelsberger2016}, which covers both WDM and DAO models, to investigate the abundance and inner structure of dark matter haloes at high redshift ($z\geq5$). \citet{Bohr2020} presented a convenient parametrization of these different structure formation models based only on two parameters $h_{\rm peak}$ and $k_{\rm peak}$, the amplitude and scale of the first DAO peak. CDM and WDM are both included in this parametrization by taking $k_{\rm peak}\rightarrow\infty$ in the former and $h_{\rm peak} = 0$ in the latter. Specifically, our objective is mainly to describe the behaviour of i) the halo mass function and ii) the halo concentration-mass relation across the ETHOS models in the simulation suite, and to interpret the results based on the Extended Press-Shechter (EPS) formalism. The latter objective is particularly relevant since it
offers an alternative to quickly compute statistical halo properties, which have so far not been fully tested across the broad range of dark matter models explored in the ETHOS framework.

We have shown that the EPS formalism within the ellipsoidal collapse model (Eqs.~\ref{eq:hmf}$-$\ref{eq:ellipsoidal}) using a smooth-$k$ window function (Eq.~\ref{eq:smooth}) with the fitting parameters $\beta=3.46$, $c_{\rm W}$=3.79 is able to accurately reproduce the halo mass function (in the redshift range $5\leq z\lesssim19$ and mass range $10^7$\msunh$\lesssim M_{200}\lesssim10^{11}$\msunh) for CDM and ETHOS models with $h_{\rm peak}=0.6-1$ (see Figs.~\ref{fig:hmf_cdm}$-$\ref{fig:hmf_z}). For models with weaker DAO features ($h_{\rm peak}<0.6$), the cut-off in the halo mass function is reproduced accurately and the overall behaviour at lower halo masses is well captured, but the accuracy below the cut-off scale is much lower than in models with higher $h_{\rm peak}$ (see Fig.~\ref{fig:hmf_kpeaks}).

Regarding halo structure, we found that the haloes of all ETHOS models at $z=5$ are well described by an NFW profile (see Fig.~\ref{fig:Q2}). The smaller the value of $k_{\rm peak}$, the lower the halo concentration towards lower halo masses relative to the CDM case.
As can be seen in Figs.~\ref{fig:concentration_cdm}$-$\ref{fig:concentration_dao}, the (median) concentration-mass relation at $z=5$ for most of the ETHOS simulations is well reproduced with the analytic model based on the EPS formalism introduced in
\citet{Ludlow2016} (tested there only for CDM and WDM).  
However, the most extreme DAO models (strong DAOs, where $h_{\rm peak}\sim1$) have a measured concentration-mass relation that behaves differently than the analytic model towards low halo masses ($M_{200}\lesssim10^9$\msunh); albeit our limited sampling of haloes (due to the small volume of our simulations) carries counting errors that remain too large to fully quantify the level of disagreement between the simulations and the analytic model. 
Simulations within a larger cosmic volume and with a larger mass range coverage are needed to firmly conclude whether an improved analytic model is needed to capture the concentration-mass relation in the full spectrum of ETHOS models.

In this work we have thus shown that it is possible to use analytic models based on the EPS formalism to reproduce the halo mass function essentially in the whole spectrum of relevant ETHOS models, that is, covering CDM, WDM and DAO models that have (allowed) galactic-scale cutoffs. This analytic prescription calibrated to our simulations has already been used in \citet{Munoz2020} to make predictions for the 21-cm hydrogen line signal during the cosmic dawn ($z\sim10-30$). We have also shown that a similar analytic approach \citep[based on][]{Ludlow2016} is able to reproduce the halo concentration-mass relation, albeit care is needed at low-masses where the reliability of the model remains unclear.

The difference between the halo mass functions across currently allowed ETHOS models will become increasingly important in the near future, when a detection/constraint in the relevant mass range becomes more feasible with upcoming observing facilities. The James Webb Space Telescope (JWST) will likely be able to probe the halo mass function indirectly through the luminosity function and test the viability of a large range of ETHOS models \citep[see e.g.][for a study of a specific wDAO model]{Lovell2018}. The hydrogen epoch-of-reionization array (HERA) will offer another promising approach to (indirectly) distinguish the different halo mass functions of ETHOS models through observation of the 21-cm signal \citep[see e.g.][for predictions based directly in the simulation suite and EPS modelling presented in this paper]{Munoz2020}.


\section*{Acknowledgements}

SB and JZ acknowledge support by a Grant of Excellence from the Icelandic Research Fund (grant number 173929). MV acknowledges support through NASA ATP grants 16-ATP16-0167, 19-ATP19-0019, 19-ATP19-0020, 19-ATP19-0167, and NSF grants AST-1814053, AST-1814259,  AST-1909831 and AST-2007355. The simulations were performed on resources provided by the Icelandic High Performance Computing Centre at the University of Iceland, and the Odyssey cluster supported by the FAS Division of Science, Research Computing Group at Harvard University.

\section*{Data availability}
The data underlying this article will be shared on reasonable request to the corresponding author.




\bibliographystyle{mnras}
\bibliography{refs} 




\appendix

\section{Halo mass function of WDM models}
\label{sec:hmf_wdm}

\begin{figure}
    \centering
    \includegraphics[width=\columnwidth]{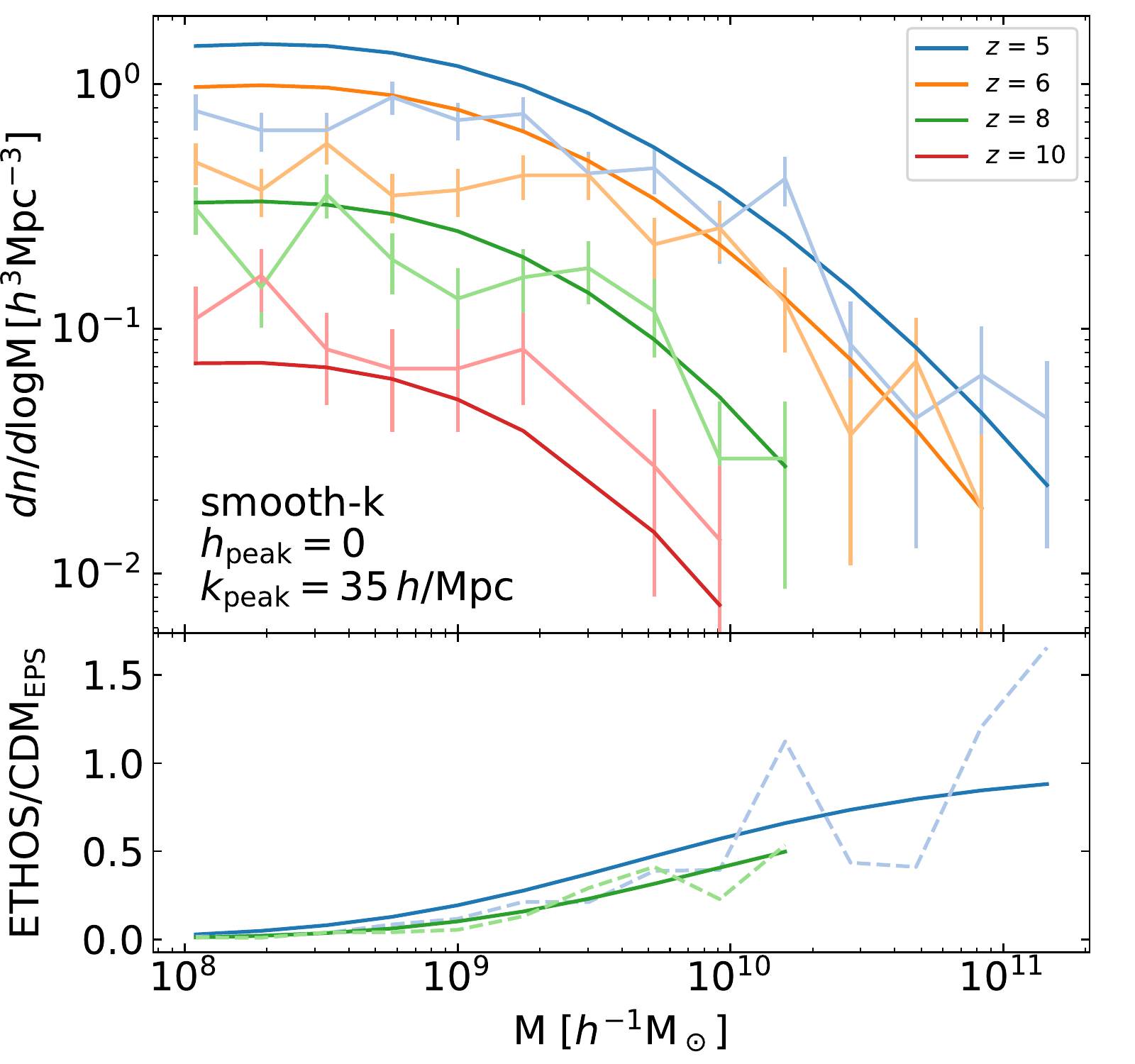}
    \includegraphics[width=\columnwidth]{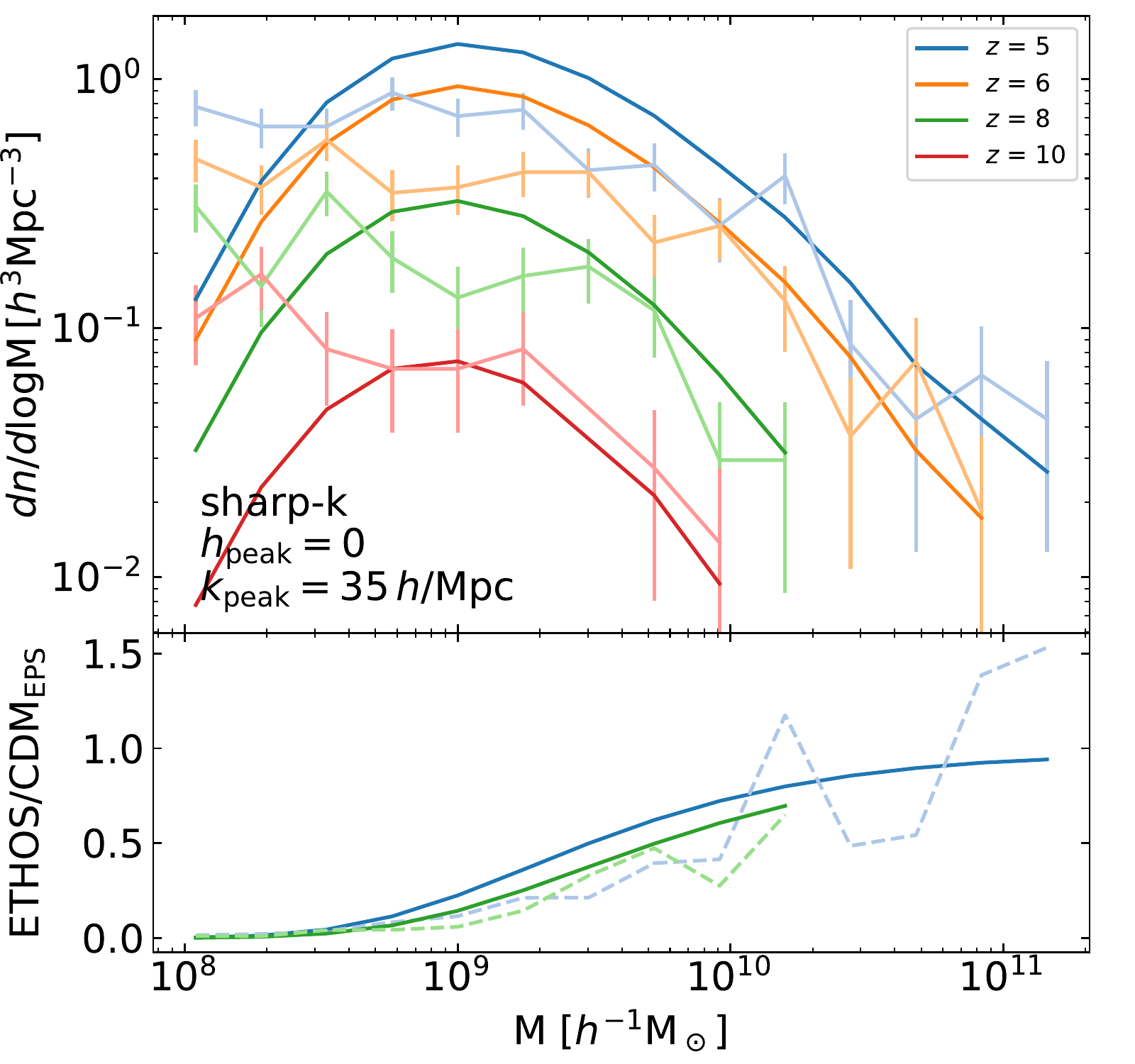}
    \caption{Halo mass function for our most extreme WDM model ($h_{\rm peak}=0$, $k_{\rm peak}=35$\hmpc) at different redshifts ($z\geq5$) according to the different colours in the legend. The light coloured lines with error bars are the simulation results, while the dark coloured lines are the EPS model using the smooth-$k$ window function (upper panels) and the sharp-$k$ window function (bottom panels). The small bottom panels for each case show the ratio of the WDM model with respect to the CDM EPS result for a selection of redshifts according to the colour legend.}
    \label{fig:hmf_wdm_z}
\end{figure}

Figure~\ref{fig:hmf_wdm_z} shows a comparison between the redshift evolution of the halo mass function of the most extreme WDM model in our simulations and the result of the EPS formalism computed with the smooth-$k$ (eq.~\ref{eq:smooth}) and the sharp-$k$ window functions. The sharp-$k$ window function is given by
\begin{equation}
    \Tilde{W}_R^{{\rm sharp-}k} (k) = \theta(1-kR),
\end{equation}
where $\theta$ is the Heaviside step function.
In the case of the sharp-$k$ window function, the sharp cut-off in the window function leads to a sharp cut-off in the halo mass function, which is clearly not a feature we resolve in our simulations before the appearance of spurious haloes for masses a few times $10^8$\msunh. Given this limitation in resolution, and given the limited sampling of haloes in our simulations for the models with the strongest cutoffs in the power spectrum (low $k_{\rm peak}$; particularly at high redshift), it is not possible to convincingly establish which of the window function achieves a better modelling of the halo mass function.

\section{Convergence test for halo density profiles}
\label{sec:convergence}

\begin{figure}
    \centering
    \includegraphics[width=\columnwidth]{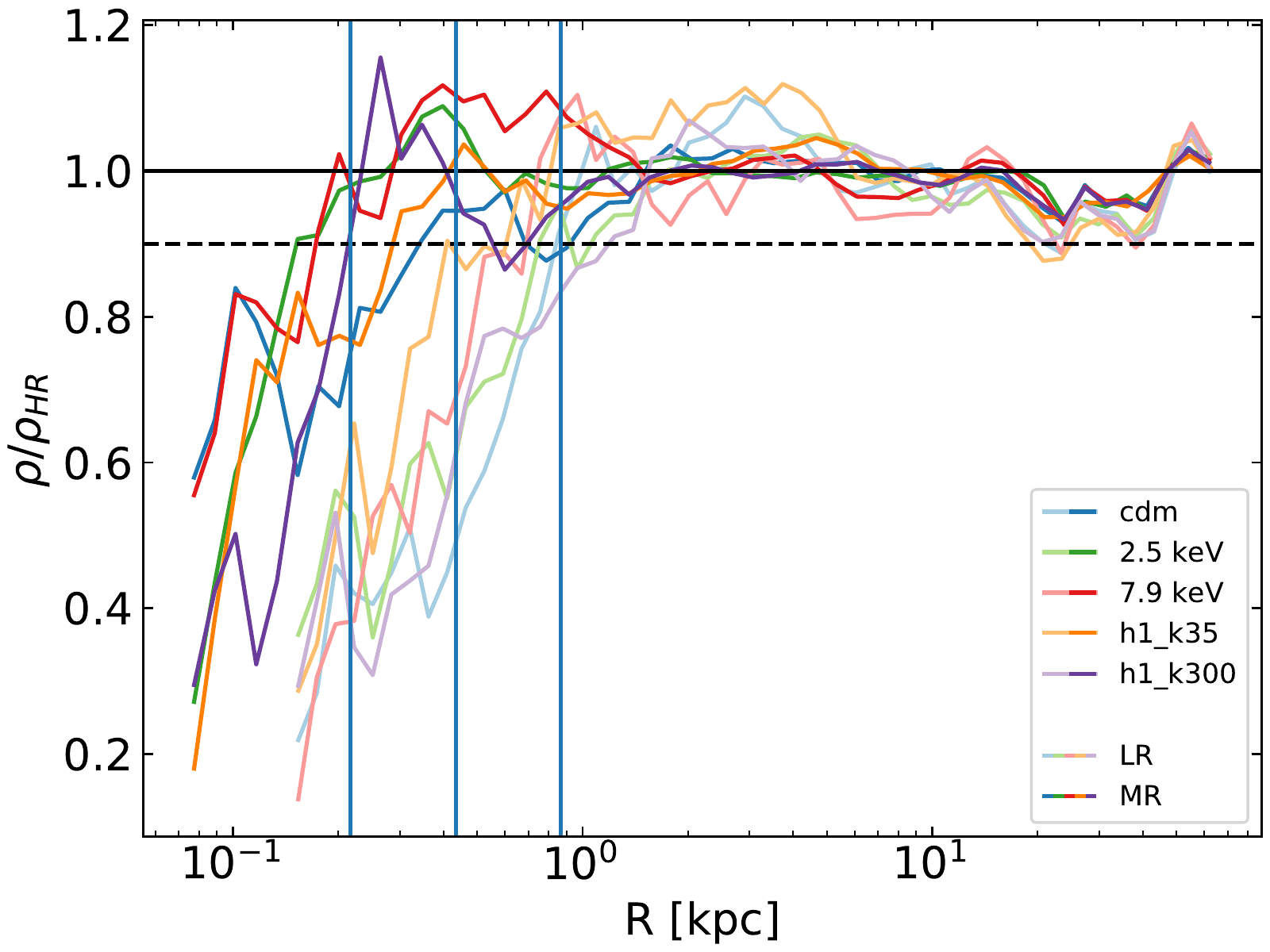}
    \caption{Convergence of the halo density for 5 models covering our parameter space. The vertical axis is the ratio of the density of the low-resolution (faded lines) and medium resolution (solid lines), relative to that of the highest resolution run at $z=5$. The dashed line indicates a convergence level of 10\%. The blue vertical lines indicate six times the softening length for high-resolution to low-resolution from left to right.}
    \label{fig:convergence}
\end{figure}

To determine the smallest radius at which we can trust the density profile measured in our simulations, we compare the density profile of the largest halo for a few DM models using three resolution levels. The chosen models cover representative regions of the parameter space and show the range of possible convergence levels. The three resolution levels were done with smoothing lengths $\epsilon=0.87$ (LR), $0.43$ (MR), and $0.22\,{\rm ckpc}/h$ (HR). As the information at the smallest scales in a halo is absent due to limited resolution, we looked for the radius below which the density in the two lower resolution levels drops by more than 10\% with respect to the highest resolution. Figure~\ref{fig:convergence} shows the ratio of the density profile between the MR and the HR levels (solid lines) and between the LR and HR levels (faded lines).
At large radii, this ratio fluctuates only slightly around 1 
and then drops substantially below six times the smoothing length, which is indicated by the vertical lines. We have therefore assigned $6\epsilon$ as the smallest resolvable scale in our high resolution simulations.

\section{Relaxation criteria for haloes}
\label{sec:relaxation}

The assembly of haloes is a very dynamic process, but the NFW profile describes a halo in equilibrium. Substantial departures from equilibrium in a halo would result in substantial deviations over the NFW profile. Therefore, we have to clean our halo catalogue by selecting only the haloes that are sufficiently virialised, have a subhalo population that is clearly subdominant by mass, and are not currently in the process of merging with a massive substructure. We adopt the relaxation criteria of \citet{Neto2007} to accomplish these goals:
\begin{itemize}
    \item The mass fraction in subhaloes must be low $f_{\rm sub}={\rm M_{sub}}/{\rm M}_{200}<0.1$, where $M_{\rm sub}$ is the total mass in subhaloes.
    \item The distance between the minimum of the potential and center of mass of the halo must be small compared to the virial radius $d_{\rm off}=|r_{\rm pot}-r_{CM}|/R_{200}<0.07$
    \item The virial ratio must be close to virialization $2|T/U|$<1.5, where $T$ and $U$ are the total kinetic and potential energies, respectively. 
\end{itemize}
We note that we have relaxed the threshold for $2|T/U|$ from 1.35 to 1.5 as we are considering a higher redshift than those studies that typically used these criteria (such as \citealt{Neto2007}). At high redshift, we expect haloes to be less virialized than at low redshift \citep[see e.g. Fig.3 in][]{Zjupa2017}.

\begin{figure}
    \centering
    \includegraphics[width=\columnwidth]{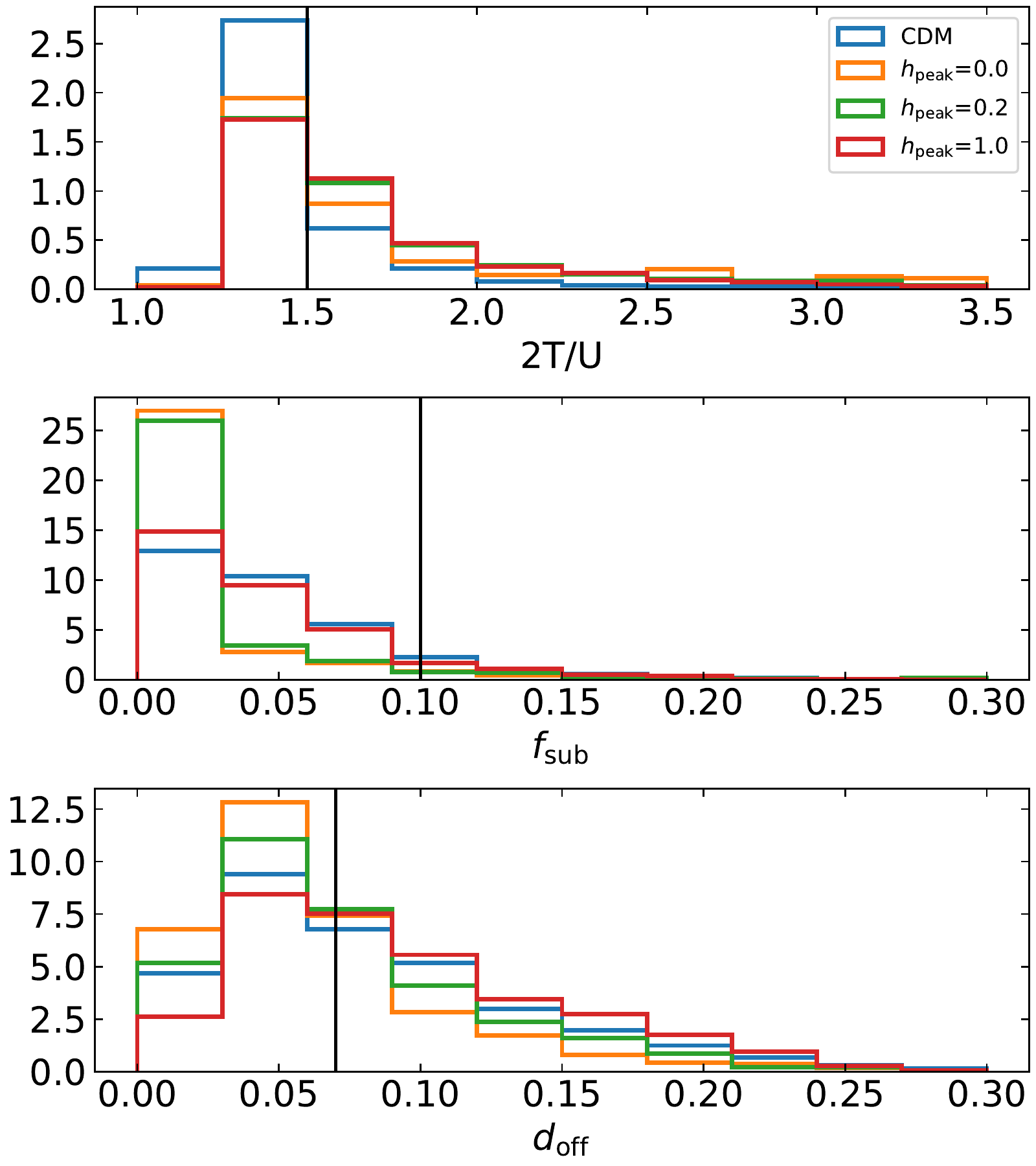}
    \caption{The distribution of the three relaxation criteria $2|T/U|$, $f_{\rm sub}$ and $d_{\rm off}$. The black vertical lines indicate our used thresholds for determining relaxed haloes. All ETHOS models have $k_{\rm peak}=35$\hmpc.}
    \label{fig:relaxation_hist}
\end{figure}

Figure~\ref{fig:relaxation_hist} shows the distribution of the three relaxation criteria for the haloes of sDAO, wDAO, and WDM models with $k_{\rm peak}=35$\hmpc, as well as CDM. We can see that a smaller number of haloes are virialized in non-CDM models than in CDM.
This trend probably arises from the delayed halo formation in models with a galactic-scale cutoff. The suppression of small scale structure in the case of WDM or wDAO is also clearly visible in the distribution of $f_{\rm sub}$. However, we find that the most restrictive criteria is given by the limit in the $d_{\rm off}$ value; clearly a substantial fraction of haloes at these redshifts are actively merging.

\bsp	
\label{lastpage}
\end{document}